\documentclass[amssymb,nobibnotes,nofootinbib,aps,prd]{revtex4}

 \usepackage{here}

\usepackage[dvipdfmx]{graphicx}
\usepackage{amssymb,amsfonts,amsmath,bm,color,multirow,cases,empheq,hyperref}
\usepackage{mathrsfs}
\definecolor{darkergreen}{rgb}{0,0.5,0}

\usepackage{enumerate}
\usepackage{ulem}
\usepackage{afterpage}

\hypersetup{%
 setpagesize=false,
 bookmarksnumbered=true,%
 bookmarksopen=true,%
 colorlinks=true,%
 linkcolor=blue,%
 citecolor=red}

\usepackage{tikz}
\usetikzlibrary{positioning,fit}
\usetikzlibrary{calc,intersections,through,backgrounds}
\usetikzlibrary{calc,arrows,decorations.markings,shapes.arrows,patterns, decorations.pathmorphing}
\usetikzlibrary{positioning,matrix,arrows,decorations.pathmorphing,decorations.pathreplacing}
\usetikzlibrary{arrows,matrix,positioning,shapes,arrows}
\usetikzlibrary{shapes.geometric}
\usetikzlibrary{decorations.text}
\usetikzlibrary{arrows.meta}

\tikzset{
  ->-/.style={decoration={markings, mark=at position 0.5 with {\arrow{to}}},
              postaction={decorate}},}
\tikzset{
  -<-/.style={decoration={markings, mark=at position 0.5 with {\arrow{to reversed}}},
              postaction={decorate}},}
              
\tikzset{
  pics/torus/.style n args={3}{
    code = {
      \providecolor{pgffillcolor}{rgb}{1,1,1}
      \begin{scope}[
          yscale=cos(#3),
          outer torus/.style = {draw,line width/.expanded={\the\dimexpr2\pgflinewidth+#2*2},line join=round},
          inner torus/.style = {draw=pgffillcolor,line width={#2*2}}
        ]
        \draw[outer torus] circle(#1);\draw[inner torus] circle(#1);
        \draw[outer torus] (180:#1) arc (180:360:#1);\draw[inner torus,line cap=round] (180:#1) arc (180:360:#1);
      \end{scope}
    }
  }
}

\newcommand{\tikznode}[2]{\relax
    \ifmmode%
    \tikz[remember picture,baseline=(#1.base),inner sep=0pt] \node (#1) {$#2$};
    \else
    \tikz[remember picture,baseline=(#1.base),inner sep=0pt] \node (#1) {#2};%
    \fi
}

\newcommand{\no}{\nonumber}

\newcommand{\cC}{\mathcal C}

\newcommand{\cL}{\mathcal L}
\newcommand{\cM}{\mathcal M}
\newcommand{\cO}{\mathcal O}

\newcommand{\la}{\lambda}
\newcommand{\Tr}{{\rm Tr}}

\allowdisplaybreaks

\begin{document}

\begin{flushright}
TTI-MATHPHYS-36
\end{flushright}
\vspace*{0.5cm}

\title{
Description of Non-Spherical Black Holes in 5D Einstein Gravity
via the Riemann-Hilbert Problem
}

\author{Jun-ichi Sakamoto}
\email{jsakamoto@toyota-ti.ac.jp}
\author{Shinya Tomizawa}
\email{tomizawa@toyota-ti.ac.jp}
\affiliation{\vspace{3mm}Mathematical Physics Laboratory, Toyota Technological Institute\vspace{2mm}\\Hisakata 2-12-1, Tempaku-ku, Nagoya, Japan 468-8511\vspace{3mm}}

\begin{abstract}
We investigate the solution-generating technique based on the Breitenlohner-Maison (BM) linear system,  for asymptotically flat, stationary, bi-axisymmetric black hole solutions with various horizon topologies in five-dimensional vacuum Einstein theory.
We construct the monodromy matrix associated with the BM linear system, which provides a unified framework for describing three distinct asymptotically flat, vacuum black hole solutions with a single angular momentum in five dimensions, each with a different horizon topology: (i) the singly rotating Myers-Perry black hole, (ii) the Emparan-Reall black ring, and (iii) the Chen-Teo rotating black lens.
Conversely, by solving the corresponding Riemann-Hilbert problem using the procedure developed by Katsimpouri et al., we demonstrate that factorization of the monodromy matrix exactly reproduces these vacuum solutions, thereby reconstructing the three geometries.
These constitute the first explicit examples in which the factorization procedure has been carried out for black holes with non-spherical horizon topologies.
In addition, we discuss how the asymptotic behavior of five-dimensional vacuum solutions at spatial infinity is reflected in the asymptotic structure of the monodromy matrix in the spectral parameter space.
\end{abstract}

\date{\today}

\maketitle

\section{Introduction}

In string theory and related fields, higher-dimensional black holes and extended objects have played a central role for a few decades. Their study has provided not only new gravitational solutions of intrinsic interest but also a testing ground for ideas that link geometry, topology, and quantum theory. 
In particular, black hole solutions of Einstein equations have long served as a fertile arena for exploring both classical and quantum aspects of gravity, ranging from questions of stability and uniqueness to deep issues of thermodynamics and information.
Over the past two decades, higher-dimensional black holes have become a major focus of investigation. 
A celebrated example is their role in the microscopic derivation of the Bekenstein-Hawking entropy~\cite{Strominger:1996sh}, which offered one of the earliest and most compelling confirmations of string theory as a candidate for a quantum theory of gravity. Another motivation has come from collider scenarios in models with large extra dimensions~\cite{Argyres:1998qn}, where black hole production at TeV scales was suggested as a possible experimental signature. These developments placed higher-dimensional black holes at the center of discussions bridging fundamental theory and potential phenomenology.
In parallel, theoretical progress has revealed a remarkable richness of the solution space. 
A broad variety of exact and approximate black hole solutions have been discovered in Einstein gravity and in various supergravity theories, often aided by modern solution-generating techniques. 
These include methods based on hidden symmetries, sigma models, and integrability structures, which have enabled the construction of families of solutions far beyond those accessible by direct integration of Einstein equations. 
Such advances have uncovered various horizon topologies, non-trivial asymptotics, and multi-centered configurations that have no analogue in four dimensions.
Despite this progress, a comprehensive classification of higher-dimensional black holes remains out of reach. 
Unlike in four dimensions - where uniqueness theorems restrict stationary, asymptotically flat black holes to the Kerr-Newman family - the higher-dimensional landscape is far richer and encodes additional degrees of freedom. 
The possible horizon topologies are not limited to spheres, 
and dynamical instabilities hint at transitions to new phases of black objects. 
These features suggest that the full structure of higher-dimensional black holes has only begun to be charted, and much remains to be found.

\medskip

For asymptotically flat, static solutions of the higher-dimensional vacuum Einstein equations, the Schwarzschild-Tangherlini solution~\cite{Tangherlini:1963bw} is the unique solution~\cite{Gibbons:2002bh,Gibbons:2002av}.
These are properties shared with the four-dimensional case.
However, the situation changes drastically for the stationary cases.
The topology theorem for stationary five-dimensional black holes~\cite{Hollands:2007aj} establishes that, under the assumptions of asymptotic flatness and the existence of two commuting axial Killing fields, the cross section of the event horizon must be either a three-sphere $S^3$, a ring $S^1\times S^2$, or a lens space $L(p,q)$. 
More generally, the horizon cross section must be of positive Yamabe type under the dominant energy condition~\cite{Galloway:2005mf,Cai:2001su}. 
 Emparan and Reall~\cite{Emparan:2001wn} first constructed the exact solution for an $S^1$-rotating black ring, thereby demonstrating that five-dimensional vacuum Einstein theory admits not only an $S^1$-rotating spherical black hole~\cite{ Myers:1986un} but also two distinct black ring solutions with identical conserved charges. This result provided a clear manifestation of the breakdown of uniqueness in higher dimensions. 
Pomeransky and Sen'kov~\cite{Pomeransky:2006bd} subsequently succeeded in constructing the balanced doubly rotating black ring solution. Although their work presented only the balanced case, an unbalanced generalization was later obtained explicitly in Ref.~\cite{Morisawa:2007di}, and a more compact form of the solution was subsequently provided in Ref.~\cite{Chen:2011jb}.
In contrast, vacuum black hole regular solutions with the horizon topology of a lens space have proven far more elusive and remain unknown. Using the inverse scattering method, several authors attempted to construct asymptotically flat solutions to the five-dimensional vacuum Einstein equations, but all such efforts ultimately failed~\cite{Evslin:2008gx,Chen:2008fa}. 
A major obstacle in obtaining a regular black lens is that the resulting solutions invariably suffer from naked singularities. Recently, however, asymptotically flat supersymmetric black lens solutions with horizon topology $L(2,1)=S^3/\mathbb{Z}_2$ and more generally $L(n,1)=S^3/\mathbb{Z}_n$ were constructed within five-dimensional minimal ungauged supergravity~\cite{Kunduri:2014kja,Breunholder:2017ubu,Tomizawa:2016kjh}. These constructions crucially rely on the powerful framework developed by Gauntlett et al.~\cite{Gauntlett:2002nw}.

\medskip

Although the uniqueness theorems for asymptotically flat, stationary, spherical black holes have been extended to both the vacuum~\cite{Morisawa:2004tc} and charged~\cite{Tomizawa:2009ua} cases in five dimensions, they still do not rule out the existence of other black hole solutions, even under the assumptions of spherical horizon topology and the spacetime symmetries of stationarity and bi-axisymmetry. This is because the topological censorship theorem (TCT)~\cite{Friedman:1993ty} suggests the existence of spherical black holes whose exterior regions may admit non-trivial topological structures.
The TCT states that, under the averaged null energy condition, the domain of outer communication (DOC) in an asymptotically flat spacetime must be simply connected. In four dimensions, this implies that the topology of the black hole exterior intersected with a spatial slice $\Sigma$ is restricted to the trivial structure ${\mathbb R}^3\setminus{\mathbb B}^3$, where ${\mathbb B}^3$ denotes the black hole region. In higher dimensions, however, the DOC can admit non-trivial topologies. 
Based on this, Ref.~\cite{Hollands:2010qy} showed that in five dimensions, ${\rm DOC}\cap\Sigma$ can have the non-trivial topology $[{\mathbb R}^4\setminus{\mathbb B}^4 \ \#\ n(S^2\times S^2)\ \#\ m(\pm{\mathbb C}P^2)]$.
In the uniqueness theorems~\cite{Morisawa:2004tc,Tomizawa:2009ua}, the exterior region of a black hole is assumed to have the trivial topology ${\mathbb R}^4\setminus{\mathbb B}^4$, with ${\mathbb B}^4$ representing the black hole region on a spatial slice $\Sigma$. 
Kunduri and Lucietti~\cite{Kunduri:2014iga} were the first to construct supersymmetric black hole solutions in five-dimensional minimal supergravity with $S^3$ horizon topology but with ${\rm DOC}$ having the non-trivial topology $[{\mathbb R}^4\ \#\ (S^2\times S^2)]\setminus{\mathbb B}^4$. 
Furthermore, Refs.~\cite{Suzuki:2023nqf,Suzuki:2024phv,Suzuki:2024abu} presented the first non-supersymmetric exact solution of an asymptotically flat, stationary spherical black hole in five-dimensional minimal supergravity, whose DOC has the topology $[{\mathbb R}^4\ \#\ {\mathbb C}P^2]\setminus{\mathbb B}^4$.
The existence of these solutions strongly suggests that many more, as yet undiscovered, black hole solutions with novel topological structures remain to be found.

When a $D$-dimensional spacetime admits $D-2$ commuting Killing vectors, Einstein gravity can be dimensionally reduced to a classically integrable two-dimensional nonlinear sigma model~\cite{Maison:1979kx}.
Belinsky and Zakharov~\cite{Belinsky:1971nt,Belinsky:1979mh} developed the inverse scattering method (ISM) for the four-dimensional vacuum Einstein equations. Their construction relies on the fact that the Einstein second-order nonlinear partial differential equations can be reformulated as a pair of first-order linear equations known as a Lax pair. This framework extends naturally to $D$-dimensional vacuum spacetimes with $(D-2)$ commuting Killing vectors.
In higher dimensions, however, a straightforward application of the ISM often produces singular spacetimes. Pomeransky refined the method, successfully deriving the five-dimensional Myers-Perry black hole from the five-dimensional Schwarzschild solution~\cite{Pomeransky:2005sj}. The ISM was also shown to generate $S^2$-rotating black rings~\cite{Tomizawa:2005wv}. Constructing the $S^1$-rotating black ring, by contrast, proved significantly more difficult, since regular seed solutions invariably lead to naked curvature singularities.
A breakthrough came in Refs.~\cite{Iguchi:2006rd,Tomizawa:2006vp}, where it was shown that choosing an appropriate singular seed solution allows one to generate the $S^1$-rotating black ring. This paved the way for the construction of the doubly rotating black ring via the ISM, and ultimately Pomeransky and Sen’kov obtained the balanced doubly rotating black ring solution~\cite{Pomeransky:2006bd}. Although their work presented only the balanced case, the unbalanced generalization was later given explicitly in Ref.~\cite{Morisawa:2007di}, and a more compact representation of the solution was subsequently provided in Ref.~\cite{Chen:2011jb}. 
More recently, by combining the ISM with Ehlers and Harrison transformations~\cite{Giusto:2007fx,Bouchareb:2007ax}, a variety of new black hole solutions and known solutions have been constructed. The Ehlers transformation~\cite{Giusto:2007fx}  generates angular momentum, while the Harrison transformation~\cite{Bouchareb:2007ax} introduces electric charge, each acting on a five-dimensional vacuum solution while preserving asymptotic flatness. These include the vacuum and charged rotating black rings~\cite{Suzuki:2024eoz,Suzuki:2024coe,Suzuki:2024vzq}, as well as a spherical black hole whose domain of outer communication (DOC) on a timeslice has the nontrivial topology $[{ \mathbb R}^4 \# { \mathbb C}{ \mathbb P}^2] \setminus { \mathbb B}^4$~\cite{Suzuki:2023nqf,Suzuki:2024phv,Suzuki:2024abu}. 
Thus, the ISM enables us to construct a wide variety of exact five-dimensional black hole solutions via the inverse scattering method~\cite{
Mishima:2005id,
Tomizawa:2005wv,Tomizawa:2006jz,Tomizawa:2006vp,
Iguchi:2006rd,
Elvang:2007rd,
Tomizawa:2007mz,Iguchi:2007xs,
Pomeransky:2006bd,
Iguchi:2007is,Evslin:2007fv,
Elvang:2007hs,Izumi:2007qx,
Chen:2011jb,
Chen:2008fa,
Chen:2012kd,
Rocha:2011vv,
Rocha:2012vs,
Chen:2015iex,
Chen:2012zb,
Lucietti:2020ltw,
Lucietti:2020phh,
Morisawa:2007di,Evslin:2008gx,
Feldman:2012vd,
Chen:2010ih,
Iguchi:2011qi,
Tomizawa:2019acu,
Tomizawa:2022qyd,
Suzuki:2023nqf,
Figueras:2009mc}.
In these works, the transformed solutions depend critically on the choice of seed solutions, where suitable (often singular) seeds were identified through a process of trial and error. 
In general, selecting appropriate seeds that yield regular solutions remains a difficult and subtle problem.
Therefore, developing techniques that circumvent the reliance on seed solutions is of crucial importance.

In this work, we aim to explore whether the solution-generating technique based on the Breitenlohner-Maison (BM) linear system \cite{Breitenlohner:1986um}---developed in \cite{Breitenlohner:1986um,Chakrabarty:2014ora,Katsimpouri:2012ky,Katsimpouri:2013wka,Katsimpouri:2014ara}---can serve as one of the unified approaches needed to treat methods such as the inverse scattering method, Ehlers transformations, and Harrison transformations within a single framework. 
A key advantage of this approach is that it does not rely on a specific choice of seed solutions. Instead, the central object is the monodromy matrix ${\cal M}(w)$ associated with the BM linear system. 
 The matrix is a meromorphic function of an auxiliary complex variable $w$, called the spectral parameter, and takes values on the Geroch group, which is an infinite-dimensional symmetry group underlying the 2D integrable coset sigma model. 
 For 5D vacuum Einstein theory, asymptotically flat, stationary and bi-axisymmetric black holes are uniquely determined by the asymptotic charges, the mass and two angular momenta and the rod data~\cite{Hollands:2007aj},  which includes the information on the topologies of the event horizon and the DOC.
Therefore,  clarifying how such rod data is encoded in the monodromy matrix would be useful, when  attempting to establish a systematic procedure to construct new black hole solutions. 
The exact gravitational solutions can be systematically constructed by solving a Riemann-Hilbert problem that involves factorizing the monodromy matrix, namely ${\cal M}(w) = {\cal V}^{\sharp}(\lambda,x){\cal V}(\lambda,x)$ ($\sharp$: anti-involution), where ${\cal V}(\lambda,x)$ is the coset element of the BM linear system, with another spectrum parameter $\lambda$ and two-dimensional coordinates $x$. 
In general, solving this factorization procedure is highly nontrivial. 
For five-dimensional non-extremal black holes with spherical horizon topology, it is known that the monodromy matrix is a matrix-valued meromorphic function with only simple poles in $w$, and the associated residues are constant matrices independent of the Weyl-Papapetrou coordinates $x=(z,\rho)$~\cite{Chakrabarty:2014ora}.
In this case, the factorization problem reduces to solving certain algebraic equations. 
Thus, an advantage of this procedure is that once an appropriate monodromy matrix is specified, the corresponding exact solution can be automatically constructed.  On the other hand, there is currently no systematic method for obtaining a monodromy matrix describing a physically acceptable black hole solution, and hence the procedure is not yet useful for generating new black hole solutions. Indeed, most previous works have focused on constructing the monodromy matrices corresponding to known exact solutions with spherical horizons, and verifying that the original solutions can be reproduced by factorization~\cite{Chakrabarty:2014ora}.

\medskip

Motivated by these considerations, the purpose of the present work is to clarify whether the solution-generating technique based on the BM linear system can be applied to the case of black holes with non-spherical horizon topology. 
In Ref.~\cite{Chakrabarty:2014ora}, the first attempt was made to construct the monodromy matrix for the Emparan-Reall black ring solution, but divergences appeared in some components of the monodromy matrix. 
This issue prevents a straightforward application of the method in that case. 
Consequently, the applicability of this technique to solutions with non-spherical horizon topology has remained unclear.
In this work, we consider the following three distinct asymptotically flat black hole solutions with a single angular momentum in five-dimensional vacuum Einstein theory--stationary and bi-axisymmetric--each characterized by a different horizon topology: a sphere $S^3$, a ring $S^1 \times S^2$, and a lens space $L(n,1)$.
\begin{itemize}
    \item the singley rotating Myers-Perry black hole \cite{Myers:1986un}
    \item the Emparan-Reall black ring \cite{Emparan:2001wn}
    \item  the Chen-Teo rotating black lens \cite{Chen:2008fa}
\end{itemize}

As in the Ehlers transformation discussed in Ref.~\cite{Giusto:2007fx}, to avoid this divergence, when constructing the monodromy matrix, we use the Killing vectors $ \partial/\partial\psi $ and $ \partial/\partial\phi $, associated with the Euler angles that parametrize the $S^3_\infty$ at spatial infinity, whose metric can  be expressed as
\begin{eqnarray}
ds^2_{\rm S^3_\infty}=\frac{r^2}{4}\left[(d\psi+\cos\theta d\phi)^2+d\theta^2+\sin^2\theta d\phi^2\right],\notag
\end{eqnarray}
with $0\leq \psi <4\pi$, $0\leq \phi<2\pi$, and $0\leq \theta \leq \pi$.

While the monodromy matrix for the $SL(3,\mathbb{R})$ Geroch group is sufficient for describing five-dimensional vacuum solutions, we instead consider monodromy matrices valued in the larger Geroch group associated with $SO(4,4)$ in order to extend the framework to other theories.
This symmetry naturally arises in the moduli space of axisymmetric solutions in five-dimensional $U(1)^3$ supergravity and provides a unified description of more general non-extremal black holes supported by three abelian gauge fields and dilaton fields.
Moreover, it can be embedded into eleven- and ten-dimensional supergravity theories, which are the low-energy effective theories of string theory, thereby allowing us to exploit powerful tools from string theory in the study of such solutions.
For these reasons, we work with monodromy matrices valued in the $SO(4,4)$ Geroch group.

With the symmetry group, we show that, for each of these solutions, the monodromy matrix $\cal M$ has only simple poles in the spectral parameter $w$, 
\begin{eqnarray*}
{\cal M}(w)=Y_{\rm flat}+\sum_i\frac{A_i}{w-w_i} 
\end{eqnarray*}
and the residue matrix $A_i$ at each pole is a certain rank-2 constant matrix. 
As a result, the existing procedure for the factorization can be applied, and we confirm that the original black hole solutions can be precisely reconstructed from the corresponding monodromy matrices.
Thus, our results indicate that the solution-generating technique based on the BM linear system is effective for black hole solutions with non-spherical horizon topology.

\medskip
In Sec.~\ref{5dsugra}, we begin with a review of 5D Einstein theory and its integrable structure. 
In Sec.~\ref{sec:mp}, we present the $SO(4,4)$-valued monodromy matrix for the 5D singly rotating Myers-Perry black hole and solve the associated factorization problem. 
In Section~\ref{sec:er}, we present the monodromy matrix for the Emparan-Reall black ring 
and perform the factorization.
In Sec.~\ref{sec:ct}, we present the monodromy matrix corresponding to the 5D rotating black lens with conical singularities, originally constructed in the work of Chen and Teo. 
Finally, Sec.~\ref{sec:dis} is devoted to discussion and concluding remarks.

\section{5D pure gravity and its integrable structure}\label{5dsugra}

In this work, we are interested in the construction of exact vacuum solutions in five-dimensional pure Einstein gravity. When restricted to bi-axisymmetric solutions, the Einstein equations can be mapped to an integrable linear system, allowing the application of solution generating techniques. In this section, we give a brief overview of a procedure for constructing solutions based on the factorization of the monodromy matrix associated with the Breitenlohner-Maison (BM) linear system developed in \cite{Katsimpouri:2012ky,Katsimpouri:2013wka,Katsimpouri:2014ara}.

\subsection{Coset space description of 5D pure Einstein gravity}

We consider asymptotically flat black hole solutions of the 5D pure Einstein gravity, whose the action is given by
\begin{align}\label{5d_sugra_action}
    S_{\rm 5D}&=\int d^5x\sqrt{-g_5}R_5\,.
\end{align}
In general, constructing exact solutions to the Einstein equations is a quite nontrivial task. However, in five-dimensional spacetime, when a gravitational solution has three commuting Killing vectors, the Einstein equations reduce to a 2D integrable linear system.
This remarkable fact enables us to employ powerful solution-generating techniques to construct exact solutions.
When we work on the coordinate system such that the metric $g_{\mu\nu}\,(\mu,\nu=0,1,\dots,4)$ at spatial infinity approaches the standard flat spacetime metric 
\begin{align}\label{asy-flat}
ds^2_{5}&=g_{\mu\nu}dx^{\mu}dx^{\nu}\simeq -dt^2+r^2\sin^2\theta\,d\tilde{\phi}^2+r^2\,\cos^2\theta\, d\tilde{\psi}^2
+dr^2+r^2d\theta^2\,,   
\end{align}
the three commuting Killing vectors are taken to be $(\partial_t,\partial_{\tilde{\phi}},\partial_{\tilde{\psi}})$. Here, the angle variable $\theta$ takes a value in $0\leq \theta<\frac{\pi}{2}$, and the ranges of the angular variables $\tilde{\phi}$ and $\tilde{\psi}$ are fixed as $\tilde \psi\sim \tilde\psi+2\pi $ and $\tilde \phi\sim \tilde\phi+2\pi $, respectively.
By performing dimensional reduction along the Killing directions and dualizing the resulting one-form fields into scalars, the Einstein-Hilbert action (\ref{5d_sugra_action}) reduces to a two-dimensional dilaton gravity theory coupled to a classically integrable two-dimensional $H\backslash G$ coset sigma model with the action
\begin{align}\label{2d-sigma}
    S_{\rm 2D}^{\rm sigma}=-2\int d\rho dz\,\sqrt{g_2}\,\rho\,g_2^{mn}\Tr(\partial_{m}MM^{-1}\,\partial_{n}MM^{-1})\,.
\end{align}
The 2D system is defined on the 2D conformal flat space
\begin{align}
    ds_2^2=e^{2\nu}(d\rho^2+dz^2)\,,
\end{align}
and the scalar moduli of the 5D metric is described by the coset matrix $M(z,\rho)$ valued in the symmetric coset space $H\backslash G$.

It is known that classical solutions of the integrable sigma models can be obtained by solving a system of linear partial differential equations 
\begin{align}
    \partial_{m}\Psi(z,\rho;w)=\cL_{m}(z,\rho;w)\Psi(z,\rho;w)\,,\qquad m=z\,,\rho\,,
\end{align}
where $\Psi$ is a $G$-valued function, and $\cL$ is a $G$-valued connection depending spectral parameter $w\in \mathbb{C}$, referred as the Lax pair. The Lax pair is specified such that the compatibility condition of the linear system, or equivalently the flatness condition of $\cL$ is equivalent to the equations of motion of the integrable coset sigma model. For the purpose of constructing black hole solutions, the following two types of the linear systems are mainly used: the Belinski-Zakharov (BZ) linear system \cite{Belinsky:1971nt,Belinsky:1979mh} and the Breitenlohner-Maison (BM) linear system \cite{Breitenlohner:1986um}. In this work, we employ the BM linear system which the space of the spectral parameter is defined on a Riemann surface with a brach cut depending on the Weyl-Papapetrou coordinates $\rho$ and $z$. For the explicit expression of the Lax pair, see for example \cite{Sakamoto:2025jtn}.

A central object in the solution generating procedure based on the BM linear system is the monodromy matrix $\cM(w)$. 
As will be explained in Sec.\,\ref{sec:bm-monodromy}, for known black hole solutions, the corresponding monodromy matrix can be constructed from the coset matrix $M(z,\rho)$.
However, there is a technical subtlety for constructing the monodromy matrices. As pointed out in \cite{Giusto:2007fx}, for 5D asymptotically flat solutions, the choice (\ref{asy-flat}) of the coordinate system leads to divergent components of the associated coset matrix at spatial infinity $r\to\infty$. Such divergences implies that the corresponding monodromy matrix has a pole at $w\to \infty$, in which case the solution generating technique based on the BM linear system cannot straightforwardly be applied. Therefore, it is necessary to take a coordinate system that avoids this issue.
To this end, following \cite{Giusto:2007fx}, we introduce new angular variables defined by
\begin{align}\label{new-angle}
   \tilde{\phi}=\frac{\phi+\psi}{2}\,,\qquad \tilde{\psi}=\frac{\phi-\psi}{2}\,.
\end{align}
By performing generalized dimensional reduction along the Killing directions $S^{1}(t)\times S^{1}(\psi)\times S^{1}(\phi)$, we find that asymptotically flat solutions in 5D Einstein gravity can be recast as classical solutions of a 2D integrable coset sigma model, subject to boundary conditions where the coset matrix approaches a constant value at spatial infinity. This coordinate system has been used in the construction of monodromy matrices for black hole solutions with spherical horizon topology \cite{Chakrabarty:2014ora}. As we will show below, it also enables the construction of well-defined monodromy matrices for black ring and lens space solutions.

\subsubsection{Parametrization of coset matrix}

As a preparation for the later discussion of solution generating techniques, we here summarize the coset matrix description of the 5D axisymmetric vacuum solutions and introduce the associated notation which mostly follows the previous work \cite{Sakamoto:2025jtn}.

As described in the previous section, by taking the angular variables $\phi$ and $\psi$, and introducing the Weyl-Papapetrou coordinates $\rho$ and $z$, the five-dimensional spacetime metric can be written as
\begin{align}
\begin{split}
    ds_5^2&=-f^2(dt+\check{A}^0)^2+f^{-1}e^{2U}(d\psi+\omega_3)^2+f^{-1}e^{-2U}(e^{2\nu}(d\rho^2+dz^2)+\rho^2d\phi^2)\,,\\
    \check{A}^0&=\zeta^{0}(d\psi+\omega_3)+\hat{A}^{0}\,,\qquad \omega_3=\omega_{3,\phi}d\phi\,,\qquad \hat{A}^{0}=\hat{A}^{0}_{\phi}d\phi\,,
\end{split}
\end{align}
where all fields only depend on $\rho$ and $z$.
We consider a dimensional reduction from five to two dimensions, following the same step as in \cite{Sakamoto:2025jtn}:
\begin{align}
    S^{1}(t)\to S^{1}(\psi)\to S^{1}(\phi)\,.
\end{align}
While the present work focuses on constructing vacuum solutions, our goal is to develop a unified, algebraic classification of more general asymptotically flat black hole solutions, including those with non-trivial Abelian gauge fields arising from Ramond-Ramond fields in string theory. To this end, we formulate the resulting two-dimensional sigma model in terms of the following symmetric coset structure:
\begin{align}\label{coset}
    \frac{G}{H}=\frac{SO(4,4)}{SO(2,2)\times SO(2,2)}\,.
\end{align}
The semisimple Lie algebra $\mathfrak{g}=\mathfrak{so}(4,4)$ is spanned by the 28 generators $\{H_{\Lambda}, E_{\Lambda},E_{q_{\Lambda}}, E_{p^{\Lambda}}\}$, and utilize the matrix representation given in the appendix of \cite{Sakamoto:2025jtn}.
The coset representative $V\in G$ in the coset space $H\backslash G$ is subject to a gauge transformation from the left by an element $h \in H$, whereas a group element $g \in G$ acts transitively from the right i.e. 
\begin{align}
    V(z,\rho)\mapsto h(z,\rho)V(z,\rho)g\,.
\end{align}
By performing a gauge transformation, we fix the coset representative in the Iwasawa gauge and parametrize it in terms of 16 scalar fields $\{\phi^a\}$ as
\begin{align}\label{iwasawa-rep}
    V&=e^{-U\,H_0}\cdot\left(\prod_{I=1}^{3}e^{-\frac{1}{2}(\log y^I)H_I}\cdot e^{-x^IE_{I}}\right)\cdot e^{-\zeta^{\Lambda}E_{q_{\Lambda}}-\tilde{\zeta}_{\Lambda}E_{p^{\Lambda}}}\cdot e^{-\frac{1}{2}\sigma E_0}\,.
\end{align}
As we will show later, nine of the sixteen scalar fields vanish for the vacuum solutions:
\begin{align}\label{vacuum-scalar}
    x^I=0\,,\qquad \zeta^I=0\,,\qquad \tilde{\zeta}_I=0\,.
\end{align}
The scalar fields $\tilde{\zeta}_0$ and $\sigma$ are obtained from the Hodge dual relations given in (\ref{dual-eq}) from the one-form fields $\check{A}^0$ and $\omega_3$.
We introduce the gauge invariant element $M(z,\rho)$ as
\begin{align}\label{m-def}
    M(z,\rho)=V^{\natural}V\,,
\end{align}
where $\natural:G\to G$ is an anti-involutive automorphism 
\begin{align}
    x^{\natural}=\eta' x^{T}\eta'\qquad 
    \eta'=\text{diag}(-1,-1,1,1,-1,-1,1,1)\qquad \text{for}\quad x\in G\,.
\end{align}
Since $\natural$ satisfies $h^{\natural} = h^{-1}$ for $h \in H$, the coset matrix $M$ is manifestly invariant under gauge transformations generated by $H$.

\subsection{BM linear system and monodromy matrix}\label{sec:bm-monodromy}

One of the central concepts in the solution-generating technique based on the Breitenlohner-Maison (BM) linear system, as developed in \cite{Breitenlohner:1986um,Chakrabarty:2014ora,Katsimpouri:2012ky,Katsimpouri:2013wka,Katsimpouri:2014ara}, is the monodromy matrix $\cM(w)$.
This matrix is a meromorphic, matrix-valued function that depends on an auxiliary complex variable $w\in \mathbb{C}$ known as the spectral parameter, and satisfies the following conditions:
\begin{align}\label{m-con}
    \cM^{-1}=\eta \cM^{T}\eta\,,\qquad   \cM^{\natural}=\cM\,.
\end{align}
At present, no general framework exists for systematically determining monodromy matrices that correspond to physically meaningful gravitational solutions. However, for specific solutions, one can construct the associated monodromy matrix by evaluating the coset matrix $M(z,\rho)$ in the limit $\rho \to 0^+$ in a region where $z$ is sufficiently negative:
\begin{align}\label{sub-rule}
    \cM(w)=\lim_{\rho \to 0^+}M(z=w,\rho)\qquad \text{for}\quad z<-R\,.
\end{align}
While a rigorous proof is still missing, it is observed from several examples that the monodromy matrices corresponding to asymptotically flat, five-dimensional non-extremal black hole solutions take the following universal expression:
\begin{align}
    \cM(w)=Y_{\rm flat}+\sum_{i=1}^{N}\frac{A_i}{w-w_i}\,.
\end{align}
Here, the constant matrix $Y_{\rm flat}=Y_{\rm flat}^{\natural
}$ characterizes the asymptotic structure of the gravitational solution and the all residue matrices $A_i$ have rank 2.
The number $N$ of simple poles expresses the number of the corner points of the rod structure (for the detail, see Refs.~\cite{Harmark:2004rm,Hollands:2007aj}), and the positions $w_i$ of simple poles are precisely identical with the locations of the corner points.
Indeed, we will see that the monodromy matrices corresponding to three vacuum solutions we consider take the same form.

Once a monodromy matrix $\cM(w)$ of this form is given, it can be factorized by following the procedure developed in \cite{Chakrabarty:2014ora,Katsimpouri:2012ky,Katsimpouri:2013wka,Katsimpouri:2014ara}, by rewriting the constant spectral parameter $w$ in terms of a coordinate-dependent spectral parameter $\lambda$ that satisfies the following algebraic relation:
\begin{align}\label{r-alg}
    \frac{1}{\la}-\la=\frac{2}{\rho}(w-z)\,,
\end{align}
Here, $w \in \mathbb{C}$ is a constant spectral parameter. In terms of $\lambda$, the monodromy matrix takes the factorized form:
\begin{align}\label{fac-m}
    \cM(w(\la,z,\rho))=X_-^{}(\la,z,\rho)M(z,\rho)X_+(\la,z,\rho)\,.
\end{align}
The matrix valued functions $X_+(\la,z,\rho)$ and $X_-(\la,z,\rho)=X_+^{\natural}(-1/\la,z,\rho)$ are normalized to satisfy the boundary conditions
\begin{align}\label{Xpm-bc}
    X_+(0,z,\rho)=1_{8\times 8}=X_-(\infty,z,\rho)\,.
\end{align}
The symbol $w(\la,z,\rho)$ in the left-hand side is to remind us that whenever $\cM(w)$ is rewritten as shown on the right-hand side, $w$ must always be substituted using its relation 
(\ref{r-alg}) with a branch 
\begin{align}\label{la-w}
    \la=\la(w;z,\rho)=\frac{1}{\rho}\left[(z-w)+ \sqrt{(z-w)^2+\rho^2}\right]\,.
\end{align}

\subsection{Asymptotic behavior of monodromy matrix for vacuum solutions}

From the relation (\ref{sub-rule}), we expect that the behavior of the monodromy matrix $\cM(w)$ in the large spectral parameter region is governed by the asymptotic behavior of the corresponding gravitational solution at spatial infinity. Here, we clarify how the asymptotic structure of asymptotically flat vacuum solutions is encoded in the algebraic structure of the corresponding monodromy matrix.

Let us consider a metric whose asymptotic behavior at spatial infinity smoothly approaches the flat spacetime metric (\ref{asy-flat}), as given below \cite{Harmark:2004rm}
\begin{align}\label{asym-met}
    ds^2_5&\sim \left(-1+\frac{8M}{3\pi}\frac{1}{r^2}+\cO\left(\frac{1}{r^3}\right)\right)dt^2-2\left(\frac{4J_{1}}{\pi }\frac{\sin^2\theta}{r^2}+\cO\left(\frac{1}{r^3}\right)\right)dtd\tilde{\phi}\no\\
    &\quad-2\left(\frac{4J_{2}}{\pi}\frac{\cos^2\theta}{r^2}+\cO\left(\frac{1}{r^3}\right)\right)dtd\tilde{\psi}\no\\
    &\quad+\left(1+\cO\left(\frac{1}{r}\right)\right)\left(dr^2+r^2(d\theta^2+\sin^2\theta d\tilde{\phi}^2+\cos^2\theta d\tilde{\psi}^2)\right)\,.
\end{align}
We write down the coset matrix $M(z,\rho)$ corresponding to this asymptotic behavior in terms of the Weyl-Papapetrou coordinates $(\rho,z)$ at infinity given by
\begin{align}
    \rho\simeq\frac{1}{4}r^2\sin2\theta\,,\qquad z\simeq \frac{1}{4}r^2\cos2\theta\,.
\end{align}
The components of the metric in the Weyl-Papapetrou coordinates behave like
\begin{align}
\begin{split}\label{metric-asy}
    g_{tt}&\simeq -1+\frac{2M}{3\pi}\frac{1}{\sqrt{\rho^2+z^2}}+\cO\left(\frac{1}{\rho^2+z^2}\right)\,,\\
    g_{t\tilde{\phi}}&\simeq -\frac{J_{1}}{2\pi}\frac{\sqrt{\rho^2+z^2}-z}{\rho^2+z^2}+\cO\left(\frac{1}{\rho^2+z^2}\right)\,,\\
    g_{t\tilde{\psi}}&\simeq -\frac{J_{2}}{2\pi }\frac{\sqrt{\rho^2+z^2}+z}{\rho^2+z^2}+\cO\left(\frac{1}{\rho^2+z^2}\right)\,,\\
    g_{\tilde{\phi}\tilde{\psi}}&\simeq \frac{\zeta}{2} \frac{\rho^2}{(\rho^2+z^2)^{\frac{3}{2}}}+\cO\left(\frac{1}{\rho^2+z^2}\right)\,,\\
    g_{\tilde{\phi}\tilde{\phi}}&\simeq 2\left(\sqrt{\rho^2+z^2}-z\right)\left(1+\frac{1}{3\pi}\frac{M+\eta}{\sqrt{\rho^2+z^2}}+\cO\left(\frac{1}{\rho^2+z^2}\right)\right)\,,\\
    g_{\tilde{\psi}\tilde{\psi}}&\simeq 2\left(\sqrt{\rho^2+z^2}+z\right)\left(1+\frac{1}{3\pi}\frac{M-\eta}{\sqrt{\rho^2+z^2}}+\cO\left(\frac{1}{\rho^2+z^2}\right)\right)\,,
\end{split}
\end{align}
where we take a limit $\sqrt{\rho^2+z^2}\to \infty$ with $z/\sqrt{\rho^2+z^2}$ fixed.
Here, the real constant $\eta$ changes under a constant shift $z\to z+\text{const.}$, and $\zeta$ is a gauge-invariant constant.
As explained in the previous section, by performing dimensional reduction to three dimensions, the asymptotic behavior of the 16 scalar fields is given by 
\begin{align}
\begin{split}\label{asym-scalar}
    e^{2U}&\simeq \sqrt{\rho^2+z^2}\left(1-\frac{\eta}{3\pi}\frac{z}{\rho^2+z^2}+\cO\left(\frac{1}{\rho^2+z^2}\right)\right)\,,\\
      x^I&=0\,,\qquad y^I\simeq1-\frac{M}{3\pi }\frac{1}{\sqrt{\rho^2+z^2}}+\cO\left(\frac{1}{\rho^2+z^2}\right)\,,\\
    \zeta^0&\simeq \frac{J_{1}}{4\pi }\frac{\sqrt{\rho^2+z^2}-z}{\rho^2+z^2}-\frac{J_{2}}{4\pi }\frac{\sqrt{\rho^2+z^2}+z}{\rho^2+z^2}+\cO\left(\frac{1}{\rho^2+z^2}\right)\,,\\
    \tilde{\zeta}_{0}&\simeq \frac{J_{1}}{4\pi }\frac{\sqrt{\rho^2+z^2}+z}{\rho^2+z^2}-\frac{J_{2}}{4\pi }\frac{\sqrt{\rho^2+z^2}-z}{\rho^2+z^2}+\cO\left(\frac{1}{\rho^2+z^2}\right)\,,\\
     \zeta^I&=0\,,\qquad \tilde{\zeta}_{I}=0\,,\\
     \sigma&\simeq 2\sqrt{\rho^2+z^2}\left(1-\frac{\eta}{3\pi}\frac{z}{\rho^2+z^2}+\cO\left(\frac{1}{\rho^2+z^2}\right)\right)\,.
\end{split}
\end{align}
If we utilize the relation (\ref{sub-rule}) to obtain the monodromy matrix, the coset matrix $M(z,\rho)$ at the spacial infinity can determine the large spectral parameter region $w\to \infty$ in the monodromy matrix $\cM(w)$.
Thus, substituting the the asymptotic behavior (\ref{asym-scalar}) of the scalar fields into (\ref{sub-rule}) leads to the asymptotic behaviour of $\cM(w)$:
\begin{align}\label{mono-asym}
    \cM(w)\simeq Y_{\rm flat}\left(1+\frac{Q}{w}\right)+\cO\left(\frac{1}{w^2}\right)\,,
\end{align}
where the asymptotic constant matrix $Y_{\text{flat}}$ is
\begin{align}\label{eq:yflat}
    Y_{\text{flat}}=
    \begin{pmatrix}
        1&0&0&0&0&0&0&0\\
        0&1&0&0&0&0&0&0\\
        0&0&0&0&0&0&0&1\\
        0&0&0&0&0&0&-1&0\\
        0&0&0&0&1&0&0&0\\
        0&0&0&0&0&1&0&0\\
        0&0&0&1&0&0&0&0\\
        0&0&-1&0&0&0&0&0
    \end{pmatrix}
    \,,\qquad 
    Y_{\text{flat}}^{\natural}=Y_{\text{flat}}
    \,.
\end{align}
By using the relation (\ref{m-con}), we can show that the matrix $Q$ defined in (\ref{mono-asym}) takes a value in $\mathfrak{so}(4,4)$, and it is expanded as 
\begin{align}\label{Q-gen}
    Q&=-\frac{M}{3\pi}\sum_{j=1}^{3}H_j
    +Q_{E_0}E_0+F_0+\frac{J_1-J_2}{2\pi}(E_{p_0}+E_{q^0})\,.
\end{align}
The coefficient of $F_0$ measures the orbifold charge acting on the $S^3$ part in the 5D asymptotically flat metric (\ref{asym-met}), and it is fixed to be one in our set up.
On the other hand, the coefficient $Q_{E_0}$ of $E_0$ is not determined by the leading asymptotic behavior (\ref{asym-scalar}) of the metric, but requires contributions from the next-order terms in $e^{2U}$ and $\sigma$.

As we will see later, the general expression (\ref{Q-gen}) for $Q$ can be verified to hold in all three examples.
This matrix is referred to as the “charge matrix” because it is directly associated with the asymptotic conserved quantities of black hole solutions.
Originally, the charge matrix was defined from the asymptotic behavior of the coset matrix $M(z,\rho)$ at spatial infinity (see, for example, \cite{Bossard:2009at,LindmanHornlund:2010gen}).
Thanks to the relation (\ref{sub-rule}) between the coset matrix and the monodromy matrix, the asymptotic behavior of the monodromy matrix is also characterized by this charge matrix.
In the $SL(3,\mathbb{R})$ case, this relation was previously discussed in \cite{Chakrabarty:2014ora}.

\subsubsection*{Positive energy theorem}

Let us consider the most general monodromy matrix with a single simple pole 
\begin{align}\label{one-mono}
    \cM_{1\text{-pole}}(w)=Y_{\rm flat}\left(1+\frac{Q_{1\text{-pole}}}{w}\right)
\end{align}
with the charge matrix (\ref{Q-gen}) describing a five-dimensional asymptotically flat spacetime. Previous studies have shown that, in the case of non-extremal black holes, the residue matrix associated with the simple pole has rank 2 \footnote{All physically admissible gravitational solutions satisfy this assumption. Indeed, a monodromy matrix associated with an extremal black hole can have the residue matrix with rank greater than two at a simple pole.}. Accordingly, we shall also assume that the residue matrix has rank 2.
Since $\text{rank}\,F_0=2$, the coefficients of other generators except for $F_0$ must be zero, and then the monodromy matrix (\ref{one-mono}) takes the expression
\begin{align}\label{one-mono2}
    \cM_{1\text{-pole}}(w)=Y_{\rm flat}\left(1+\frac{F_0}{w}\right)=Y_{\rm flat}\exp\left(\frac{1}{w}F_0\right)\,,
\end{align}
where in the second equality we used the nilpotent property $F_0^2=0$.
Since the ADM mass $M=0$ vanishes, the positive energy theorem \cite{Witten:1981mf,Chrusciel:2005pb} implies that the spacetime described by the monodromy matrix (\ref{one-mono2}) is expected to be five-dimensional Minkowski spacetime.
However, it is noted that at this stage neither the angular momenta $J_{1,2}$ nor $Q_{E_0}$ are necessarily zero from the general form (\ref{Q-gen}) of the charge matrix.
To perform the factorization, the constant spectral parameter $w$ needs to be expressed in terms of $\la, z, \rho$ using (\ref{la-w}).
By denoting $\la_0$ by $\la(w=0;z, \rho)$ in (\ref{la-w}), the inverse of $w$ is expressed as
\begin{align}\label{w-la-map}
    \frac{1}{w}=\nu_0\left( \frac{\la_0}{\la-\la_0}+\frac{1}{1+\la \la_0}\right)\,,\qquad 
     \nu_0=-\frac{2}{\rho\left(\la_0+\la_0^{-1}\right)}\,.
\end{align}
With simple algebraic manipulations, the monodromy matrix can be rewritten in a factorized form:
\begin{align}
    \cM_{1\text{-pole}}(w)&=Y_{\rm flat}\exp\left(\frac{1}{w}F_0\right)\no\\
    &=Y_{\rm flat}\exp\left(\nu_0\left( \frac{\la_0}{\la-\la_0}+\frac{1}{1+\la \la_0}\right)F_0\right)\no\\
    &=\exp\left(\frac{\nu_0\la_0}{\la- \la_0}F_0^{\natural}\right)\Bigl[Y_{\rm flat}\exp\left(\nu_0F_0\right)\Bigr]\exp\left( -\frac{\nu_0\la \la_0}{1+\la \la_0}F_0\right)\no\\
    &= X_-  M_{1\text{-pole}}(z,\rho)X_+\,,
\end{align}
where the coset matrix is
\begin{align}\label{flat-mat}
    M_{1\text{-pole}}(z,\rho)=Y_{\rm flat}\exp\left(-\nu_0F_0\right)\,,
\end{align}
and the matrices $X_{\pm}$ are given by
\begin{align}
    X_+&=\exp\left( -\frac{\nu_0\la \la_0}{1+\la \la_0}F_0\right)\,,\qquad
    X_-=\exp\left(\frac{\nu_0\la_0}{\la- \la_0}F_0^{\natural}\right)\,.
\end{align}
By extracting the scalar fields from (\ref{flat-mat}) through the parametrization (\ref{iwasawa-rep}), we obtain non-trivial scalar fields
\begin{align}
    e^{2U}&= \sqrt{\rho^2+z^2}\,,\qquad
     \sigma= 2\sqrt{\rho^2+z^2}\,,
\end{align}
and find that these scalar fields precisely describe the five-dimensional Minkowski spacetime.

From the above discussion, we have shown that, given the asymptotic structure and the number of poles (the rod structure) of the corresponding spacetime, the monodromy matrix is uniquely determined under the assumption of a constraint on the rank of the residue matrices. Moreover, if we can construct the monodromy matrix uniquely from geometric data associated with gravitational solutions with more intricate rod structures, it would provide a powerful method for generating gravitational solutions. We intend to continue reporting progress in this direction.

\newpage

\section{5D Myers-Perry black hole}\label{sec:mp}

As a first illustrative example, we analyze the factorization of the monodromy matrix associated with the 5D Myers-Perry black hole with a single angular momentum, whose horizon cross section exhibits the topology of a three-sphere $S^3$. Although the monodromy matrix for the general doubly rotating Myers-Perry solution was first constructed in Ref.~\cite{Chakrabarty:2014ora} within the $SL(3,\mathbb{R})$ Geroch group, here we present the explicit form of the corresponding monodromy matrix in the larger $SO(4,4)$ Geroch group, for future reference. We subsequently solve the corresponding Riemann-Hilbert problem.

\begin{figure}
\begin{center}
\begin{tikzpicture}[scale=0.65]
\node[font=\small ] at (-10,3) {$t$};
\node[font=\small ] at (-10,2) {$\tilde{\phi}$};
\node[font=\small ] at (-10,1) {$\tilde{\psi}$};
\node[font=\small ] at (-10,0) {$z$};
\node[font=\small ] at (-6.5,1.5) {$(0,0,1)$};
\node[font=\small ] at (6.5,2.5) {$(0,1,0)$};
\node[font=\small ] at (0,3.5) {$(1,\omega_{\tilde{\phi}},0)$};
\node[font=\small ] at (-4,-0.5) {$w_1$};
\node[font=\small ] at (4,-0.5) {$w_2$};
\draw[gray,line width = 0.8] (-9,3) -- (9,3);
\draw[black,line width = 5] (-4,3) -- (4,3);
\draw[gray,line width = 0.8] (-9,3) -- (9,3);
\draw[gray,line width = 0.8] (-9,1) -- (9,1);
\draw[black,line width = 5] (4,2) -- (8.4,2);
\draw[black,line width = 5,dashed ] (8.5,2) -- (9,2);
\draw[gray,line width = 0.8] (-9,2) -- (9,2);
\draw[black,line width = 5,dashed ] (-9,1) -- (-8.5,1);
\draw[black,line width = 5] (-8.4,1) -- (-4,1);
\draw[black,dashed ] (-4,0) -- (-4,3);
\draw[black,dashed ] (4,0) -- (4,3);
\draw[->,black,line width = 1] (-9,0) -- (9,0);
\end{tikzpicture}
\caption{Rod diagram for 5D Myers-Perry solution. The positions of the intersection points of rod vectors are $w_1=-\frac{1}{2}\alpha\,, w_2=\frac{1}{2}\alpha$ with $\alpha>0$. }\label{mp-rod}
\end{center}
\end{figure}
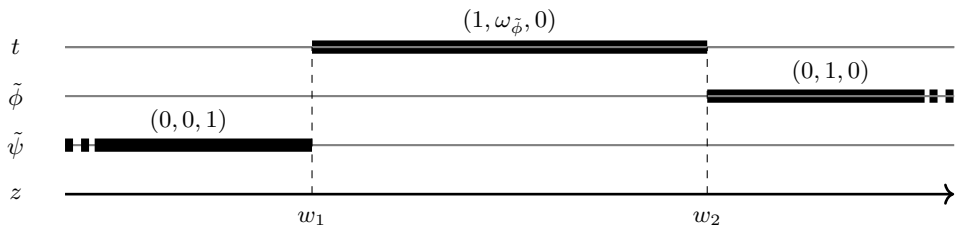

\subsection{5D Myers-Perry solution}

The metric for the 5D singly rotating Myers-Perry black hole solution  is given by
\begin{align}\label{mr-metric}
ds^2_{\rm MP}&=-dt^2+\frac{r_0^2}{\Sigma}\left[dt-a\,\sin^2\theta d\tilde{\phi}\right]^2\no\\
&\quad+(r^2+a^2)\sin^2\theta\,d\tilde{\phi}^2+r^2\,\cos^2\theta\, d\tilde{\psi}^2
+\frac{\Sigma}{\Delta}dr^2+\Sigma d\theta^2\,,    
\end{align}
with $r_0$ and $a$ denoting the mass and  rotation parameters. The  functions $\Delta$ and $\Sigma$ are defined by
\begin{align}
    \Delta&=r^2-r_0^2+a^2\,,\qquad
    \Sigma=r^2+a^2\cos^2\theta\,.
\end{align}
The angular coordinates range over
\begin{align}
   0\leq \theta<\frac{\pi}{2}\,,\qquad  0\leq \tilde{\phi}<2\pi\,,\qquad 0\leq \tilde{\psi}<2\pi\,.
\end{align}
The asymptotic conserved charges are  
\begin{align}
\begin{split}\label{mp-asymq}
    M&=\frac{3\pi}{8}r_0^2\,,\qquad J_1=\frac{\pi}{4}a r_0^2\,,\qquad J_2=0\,.
\end{split}
\end{align}
In this geometry, the Weyl-Papapetrou  coordinates $(\rho,z)$ are given by
\begin{align}
    \rho=\frac{1}{4}r \sqrt{\Delta} \sin2\theta\,,\qquad z=\frac{1}{4}r^2\left(1-\frac{r_0^2-a^2}{2r^2}\right)\cos2\theta\,.
\end{align}
with
\begin{align}
    \alpha=\frac{r_0^2-a^2}{4}\,.
\end{align}
The rod structure is depicted in Fig.\,\ref{mp-rod}, with two turning points,
\begin{align}
     w_1=-\frac{1}{2}\alpha\,,\qquad w_2=\frac{1}{2}\alpha\,,
\end{align}
which consists of three rods: 
(i) the $\tilde \psi$-rotational axis: $I_1=\{(\rho,z)|\rho=0,-\infty<z<w_1\}$, (ii) the horizon cross section: $I_2=\{(\rho,z)|\rho=0,w_1<z<w_2\}$,  (iii) the $\tilde \phi$-rotational axis: $I_3=\{(\rho,z)|\rho=0,w_2<z<\infty\}$.
The rod vector on the finite interval $I_2$ takes the form $v_2=(1,\omega_{\tilde{\phi}},0)$ with the angular velocity of the horizon, $\omega_{\tilde{\phi}}=\frac{r_0^2+a^2-4\alpha}{2a r_0^2}$. 
The rod vectors $v_1=(0,0,1)$ and $v_3=(0,1,0)$ on the semi-infinite rods indicate that $I_1$ and $I_3$ correspond to the fixed-point sets of the $U(1)$ isometries generated by $\partial_{\tilde{\psi}}$ and $\partial_{\tilde{\phi}}$, respectively.

\subsection{Coset space description}

To obtain the corresponding monodromy matrix, we first perform a dimensional reduction to three dimensions and extract the 16 scalar fields that parametrize the coset matrix $M_{\rm MP}(z,\rho)\in SO(4,4)$ with the angle variables $\phi$ and $\psi$ defined in (\ref{new-angle}), as described in the previous section.
By performing a reduction of the action (\ref{5d_sugra_action}) with the solution (\ref{mr-metric}), the resulting 16 scalar fields are given by
\begin{align}\label{mr-scalar}
\begin{split}
 e^{2U}&=\frac{r_0^2-a^2}{8}\sqrt{\frac{a^2(x-y)-r_0^2(1-y)}{a^2(x-y)+r_0^2(1+y)}}\frac{a^2(x^2-y^2)-r_0^2(1-y^2)}{a^2(x-y)-r_0^2(1-y)}\,,\\
    x^I&=0\,,\quad y^I=\sqrt{\frac{a^2(x-y)-r_0^2(1-y)}{a^2(x-y)+r_0^2(1+y)}}\,,\\
   \zeta^0&=\frac{ar_0^2(1-x)}{2(a^2(x-y)-r_0^2(1-y))}\,,\quad
   \tilde{\zeta}_{0}=\frac{ar_0^2(1+x)}{2(a^2(x-y)+r_0^2(1+y))}\,,\\
   \zeta^I&=0\,,\qquad
\tilde{\zeta}_{I}=0\,,\\
\sigma&=\frac{1}{4}(r_0^2-a^2)\biggl(y-\frac{a^2r_0^2(1-x^2)(a^2(x-y)+r_0^2y)}{(r^2_0-a^2)(a^2(x-y)-r_0^2(1-y))(a^2(x-y)+r_0^2(1+y))}\biggr)-\frac{a^2x}{4}\,.
\end{split}
\end{align}
To simply the expressions of the scalar fields, we introduced the the $C$-metric coordinates $x$ and $y$ as
\begin{align}
    x=\cos2\theta\,,\qquad y=\frac{2r^2}{r_0^2-a^2}-1\,,\qquad \,,
\end{align}
where the coordinates take values
\begin{align}
    -1\leq x \leq 1\,,\qquad  y\geq 1\,.
\end{align}
The conformal factor $e^{2\nu}$ is given by
\begin{align}\label{conf-mp}
    e^{2\nu}=\frac{r_0^2(1-y^2)-a^2(x^2-y^2)}{4\alpha\,(x^2-y^2)}\,.
\end{align}
By substituting the scalar fields (\ref{mr-scalar}) into the group element (\ref{iwasawa-rep}), we obtain the coset matrix $M_{\rm MP}(z,\rho)$ and we can see that $M_{\rm MP}(z,\rho)$ approaches the following constant matrix at the spacial infinity $r\to\infty$:
\begin{align}\label{M-mp-y}
    \lim_{r\to \infty}M_{\rm MP}(z,\rho)= Y_{\text{flat}}\,.
\end{align}
In general, the twist potentials $\tilde{\zeta}_{\Lambda}$ and $\sigma$ are defined only up to constant shifts. Here, we fix the gauge by requiring that $M_{\rm MP}(z,\rho)$ obeys the boundary condition (\ref{M-mp-y}) together with (\ref{eq:yflat}).

\subsection{Monodromy matrix}

We now compute the monodromy matrix $\cM_{\rm MP}(w)$ for the 5D Myers-Perry black hole.
According to the relation (\ref{sub-rule}) between the monodromy matrix $\cM_{\rm MP}(w)$ and the coset matrix $M_{\rm MP}(z,\rho)$, the corresponding monodromy matrix can be obtained by taking the limit $\rho\to 0$ in the region where $z$ is sufficiently negative.
The monodromy matrix $\cM_{\rm MP}(w)$ can take the form
\begin{align}\label{mr-mono}
    \cM_{\rm MP}(w)=Y_{\rm flat}+\sum_{i=1}^{2}\frac{A_i}{w-w_i}\,,
\end{align}
where the positions of the poles are
The explicit expressions of the residue matrices $A_j$ are given by
\begin{align}
\begin{split}\label{mp-A}
     A_1&=
    \begin{pmatrix}
       -\frac{r_0^4}{4(r_0^2-a^2)}&0&0&-\frac{ar_0^2}{2(r_0^2-a^2)}&0&0&\frac{a r_0^4}{16(r_0^2-a^2)}&0\\
        0&0&0&0&0&0&0&0\\
      0&0&-1&0&0&0&0&-\frac{r_0^2}{8}\\
      \frac{a r_0^2}{2(r_0^2-a^2)}&0&0&\frac{a^2}{r^2_0-a^2}&0&0&-\frac{a^2r_0^2}{8(r^2_0-a^2)}&0\\
      0&0&0&0&0&0&0&0\\
        0&0&0&0&0&0&0&0\\
       \frac{ar_0^4}{16(r_0^2-a^2)}&0&0&\frac{a^2r_0^2}{8(r_0^2-a^2)}&0&0&-\frac{a^2r_0^4}{64(r_0^2-a^2)}&0\\
       0&0&\frac{r_0^2}{8}&0&0&0&0&\frac{r_0^4}{64}\\
    \end{pmatrix}\,,\\
     A_2&=\begin{pmatrix}
       \frac{a^2r_0^2}{4(r_0^2-a^2)}&0&0&\frac{ar_0^2}{2(r_0^2-a^2)}&0&0&\frac{(r_0^2-2a^2)a r_0^2}{16(r_0^2-a^2)}&0\\
        0&0&0&0&0&0&0&0\\
      0&0&0&0&0&0&0&0\\
      -\frac{a r_0^2}{2(r_0^2-a^2)}&0&0&-\frac{r_0^2}{r^2_0-a^2}&0&0&-\frac{r_0^2(r_0^2-2a^2)}{8(r^2_0-a^2)}&0\\
      0&0&0&0&\frac{r_0^2}{4}&0&0&-\frac{a r_0^2}{8}\\
        0&0&0&0&0&0&0&0\\
       \frac{(r_0^2-2a^2)ar_0^2}{16(r_0^2-a^2)}&0&0&\frac{(r_0^2-2a^2)r_0^2}{8(r_0^2-a^2)}&0&0&\frac{(r_0^2-2a^2)r_0^2}{64(r_0^2-a^2)}&0\\
       0&0&0&0&\frac{a r_0^2}{8}&0&0&-\frac{a^2 r_0^2}{16}\\
    \end{pmatrix}\,.
\end{split}
\end{align}
While in the $SL(3,\mathbb{R})$ case the residue matrices have rank 1 \cite{Chakrabarty:2014ora}, in the $SO(4,4)$ case both residue matrices $A_j$ are of rank 2.
As observed in many examples, the poles of the monodromy matrix are located precisely at the turning points of the rods, as illustrated in Fig.~\ref{mp-rod}.

\subsubsection*{Charge matrix}

Here, we compute the $SO(4,4)$ charge matrix $Q$ introduced in the previous section.
From the expressions (\ref{mp-A}) of $A_j$, $Q$ can be expanded as
\begin{align}
    Q&=-\frac{1}{8}r_0^2 \sum_{j=1}^{3}H_j
    +\frac{1}{64}(r_0^2-4a^2)r_0^2E_0+F_0+\frac{a r_0^2}{8}(E_{p_0}+E_{q^0})\,.
\end{align}
We can confirm that this expression matches the universal form (\ref{Q-gen}) of $Q$ by using the asymptotic quantities (\ref{mp-asymq}).
We also find that $Q$ satisfies the cubic relation
\begin{align}\label{mp-q-rel}
    Q^3-\frac{1}{4}\Tr(Q^2)Q=0\,,
\end{align}
where
\begin{align}
    \Tr(Q^2)=\frac{1}{4}(r_0^2-a^2)r_0^2\,.
\end{align}
It is noted that when $r_0$ and $a$ satisfy either of the following conditions,
\begin{align}
    (\text{i})\,r_0=0\,,\qquad (\text{ii})\,r_0^2=a^2\,,
\end{align}
the charge matrix $Q$ becomes nilpotent of degree three.
Interestingly, the latter condition corresponds to the extremal limit of the five dimensional Myers-Perry black hole\footnote{The first condition corresponds to the 5D Minkowski spacetime.}. This is one of the nice properties of the charge matrix, and the extremal limit of black hole solutions are classified through the algebraic structure of its nilpotency \cite{Bossard:2009at}. 

Here, we also comment on how the charge matrix in the $SO(4,4)$ case differs from that in the $SL(3,\mathbb{R})$ case \cite{Chakrabarty:2014ora}.
Whereas the $SL(3,\mathbb{R})$ charge matrix does not encode angular momentum, the $SO(4,4)$ charge matrix already incorporates it, at least in the singly rotating case.
As a consequence, the extremal condition for the corresponding black hole solutions can be characterized directly by imposing the nilpotent condition on the $SO(4,4)$ charge matrix $Q$, rather than on its $SL(3,\mathbb{R})$ counterpart.
This is simpler than in the $SL(3,\mathbb{R})$ framework, where one must instead consider the sum of $Q$ with an additional matrix containing the angular momentum parameter.
This may be regarded as one of the advantages of formulating the monodromy matrix in the $SO(4,4)$ Geroch group rather than in $SL(3,\mathbb{R})$.

\subsection{Factorization of monodromy matrix}

We explicitly perform the factorization of the monodromy matrix (\ref{mr-mono}) by following \cite{Katsimpouri:2013wka} (see also \cite{Sakamoto:2025jtn}).
To this end, we express the residue matrices $A_j$ in terms of the eight-component vectors $a_j$ and $b_j$
\begin{align}\label{rank2-mat}
    A_j=\alpha_j(a_j\otimes a_j)\eta'-\beta_j((\eta b_j)\otimes (\eta b_j))\eta'\,,
\end{align}
where $\alpha_j$ and $\beta_{j}$ are constants.
The constant vectors $a_j$ and $b_j$ satisfy
\begin{align}\label{ab-cond}
    a_j^{T}\eta a_j=0\,,\qquad  b_j^{T}\eta b_j=0\,,\qquad a_j^{T}b_j=0\,.
\end{align}
Here, we construct these matrices $A_j$ using the eigenvectors with the non-zero eigenvalues; the explicit expressions are shown as
\begin{align}
\begin{split}
    a_1^{T}&=\left(\frac{-4r_0^2}{\sqrt{r_0^2-a^2}},0,0,\frac{8a}{\sqrt{r_0^2-a^2}},0,0,\frac{ar_0^2}{\sqrt{r_0^2-a^2}},0\right)\,,\\
    b_1^{T}&=\left(0,0,-8,0,0,0,0,r_0^2\right)\eta\,,\\
    a_2^{T}&=\left(\frac{-4 ar_0}{\sqrt{r_0^2-a^2}},0,0,\frac{8r_0}{\sqrt{r_0^2-a^2}},0,0,\frac{r_0(2a^2-r_0^2)}{\sqrt{r_0^2-a^2}},0\right)\,,\\
    b_2^{T}&=\left(0,0,0,0,2r_0,0,0,a r_0\right)\eta\,,
\end{split}
\end{align}
and the constants $\alpha_j$ and $\beta_j$ are given by
\begin{align}
    \alpha_1&=\frac{1}{64}\,,\qquad 
    \beta_1=-\frac{1}{64}\,,\qquad
    \alpha_2=-\frac{1}{64}\,,\qquad \beta_2=\frac{1}{16}\,.
\end{align}
In order to construct the matrix-valued function $X_+(\la,z,\rho)$ in the factorized form (\ref{fac-m}), we take the following ansatz such that it consists only of simple poles at $\la=\bar{\la}_{j}=-1/\la_{j}$ for all $j$ \cite{Katsimpouri:2013wka}:
\begin{align}\label{xp-ex}
    X_+(\la,z,\rho)=1-\sum_{j=1}^{N}\frac{\la C_j}{1+\la \la_j}\,,
\end{align}
where each residue $C_j$ is defined as
\begin{align}\label{c-def}
    C_j=(c_j\otimes a_j)\eta'-\left((\eta d_j)\otimes (\eta b_j)\right)\eta'\,.
\end{align}
The vectors $c_j$ and $d_j$ are obtained by the relations \cite{Katsimpouri:2013wka,Sakamoto:2025jtn}
\begin{align}\label{abcd-eq2}
    \eta'a&=d\,\Gamma^{(0)T}-(\eta c) \Gamma^{(a)T}\,,\qquad \eta'b=c\Gamma^{(0)}+(\eta d)\,\Gamma^{(b)T}\,,
\end{align}
where the $8\times N$ matrices $a,b,c,d$ are
\begin{align}
    a&=(a_1,\dots,a_N)\,,\quad  b=(b_1,\dots,b_N)\,,\quad c=(c_1,\dots,c_N)\,,\quad  d=(d_1,\dots,d_N)\,.
\end{align}
The $2\times 2$ matrices $\Gamma^{(0)}$ and $\Gamma^{(a)}\,, \Gamma^{(b)}$ are expressed in terms of the vectors $a_j,b_j$. Their precise definitions can be found in Sec.\,3 in \cite{Sakamoto:2025jtn}.
The explicit expressions of these matrices for the Myers-Perry black hole are given by
\begin{align}
    \Gamma^{(0)}&=\frac{1}{\sqrt{r_0^2-a^2}}
    \begin{pmatrix}
        \frac{64 a}{\la_1\nu_1}&-8r_0(r_0^2-a^2)\frac{1}{\la_{1,2}}\\
        -16r_0(r_0^2-a^2)\frac{1}{\la_{1,2}}&\frac{32 a}{\la_2\nu_2}
    \end{pmatrix}\,,\\
    \Gamma^{(a)}&=\Gamma^{(b)}=0_{2\times 2}\,.
\end{align}
In contrast to the static case, the matrix $\Gamma^{(0)}$ for the rotating case has non-zero diagonal components, which is proportional to the rotating parameter $a$.
Then, we obtain the matrix $X_+$ and it follows that the monodromy matrix $\cM_{\rm MP}(w)$ can be factorized
\begin{align}\label{mon-fac-mr}
    \cM_{\rm MP}(w(\la,z,\rho))=X_-(\la,z,\rho)M_{\rm MP}(z,\rho)X_+(\la,z,\rho)\,.
\end{align}
Thus, the monodromy matrix (\ref{mr-mono}) can describe the Myers-Perry black hole solution.

\subsubsection*{Conformal factor}

Finally, we compute the conformal factor $e^{2\nu}$.
Since $\Gamma^{(a)}=\Gamma^{(b)}=0_{3\times 3}$, the conformal factor $e^{2\nu}$ can be obtained by using the simplified formula \cite{Katsimpouri:2013wka} (see also \cite{Katsimpouri:2012ky})
\begin{align}\label{conf}
     e^{2\nu}&=k_{\rm BM}\prod_{j=1}^{3}(\la_j\nu_j)\,{\rm det}(\Gamma^{(0)})\,,
\end{align}
where $k_{\rm BM}$ is the integration constant and $\nu_j$ is defined as
\begin{align}
      \nu_j=-\frac{2}{\rho\left(\la_j+\la_j^{-1}\right)}\,.
\end{align}
From the expression of $\Gamma^{(0)}$, the right-hand side can be computed as
\begin{align}\label{conf-mr-com}
   &k_{\rm BM}\prod_{j=1}^{3}(\la_j\nu_j)\,{\rm det}(\Gamma^{(0)})
    =-2048k_{\rm BM}\frac{r_0^2(1-y^2)-a^2(x^2-y^2)}{4\alpha\,(x^2-y^2)}\,.
\end{align}
This precisely leads to the conformal factor (\ref{conf-mp}) for the 5D Myers-Perry black hole by taking the overall constant $k_{\rm BM}$ as
\begin{align}
    &k_{\rm BM}^{-1}=-2048\,.
\end{align}

\newpage

\section{ Emparan-Reall black ring}\label{sec:er}

Next we present the explicit expression of the monodromy matrix corresponding to the Emparan-Reall black ring solution~\cite{Emparan:2001wn} which rotates only along the $S^1$ direction  in 5D vacuum Einstein theory.
This solution was originally obtained as the Wick rotation of the solution in Ref.~\cite{Chamblin:1996kw}.
It was later reconstructed using the inverse scattering method in Ref.~\cite{Tomizawa:2006vp} and through the B\"acklund transformation in Ref.~\cite{Iguchi:2006rd}.
Here, we show that factorizing the monodromy matrix can also reproduce the Emparan-Reall black ring solution.

\subsection{Emparan-Reall black ring solution}

The metric for the the Emparan-Reall black ring solution~ \cite{Emparan:2001wn} is written in the $C$-metric coordinates $(u,v)$  as~\cite{Harmark:2004rm}
\begin{align}\label{er-bring}
    ds_5^2&=-\frac{F(v)}{F(u)}\left(dt-\cC \tilde{\kappa} \frac{1+v}{F(v)}d\tilde{\phi}\right)^2+\frac{2\tilde{\kappa}^2F(u)}{(u-v)^2}
    \biggl[-\frac{G(v)}{F(v)}d\tilde{\phi}^2+\frac{G(u)}{F(u)}d\tilde{\psi}^2
    +\frac{du^2}{G(u)}-\frac{dv^2}{G(v)}\biggr]\,,
\end{align}
where $\tilde{\kappa}$ is a real parameter, and the coordinates $u$ and $v$ take values in the ranges
\begin{align}
    -1\leq u \leq 1\,,\quad -\infty < v \leq -1\,.
\end{align}
The functions $F(x)\,, G(x)$ and the parameter $\cC$ are defined as
\begin{align}
    F(x)&=1+b x\,,\qquad G(x)=(1-x^2)(1+c x)\,,\\
    \cC&=\sqrt{2b(b-c)\frac{1+b}{1-b}}
\end{align}
with the real parameters $c$ and $b$ satisfying
\begin{align}
    0< c\leq b<1\,.
\end{align}
The event horizon is located at $v=-1/c$ and its topology is a ring $S^1\times S^2$.

In general, the metric (\ref{er-bring}) exhibits conical singularities along the rotational axes $u=-1$, $v=-1$, and $u=1$.
The singularities at $u=-1$ and $v=-1$ can be removed by fixing the periodicities of the angular coordinates $\tilde{\phi}$ and $\tilde{\psi}$ as
\begin{align}\label{psi-phi-peri}
\Delta \tilde{\psi}=2\pi\frac{\sqrt{1-b}}{1-c}\,,\qquad \Delta \tilde{\phi}=2\pi\frac{\sqrt{1-b}}{1-c}.
\end{align}
Meanwhile, the conical singularity at $u=1$ can be eliminated by imposing a constraint on the parameters $b$ and $c$:
\begin{align}\label{b-c-reg}
b=\frac{2c}{1+c^2}.
\end{align}
From  the asymptotic  form of the metric (\ref{er-bring}) at $(u,v)=(-1,-1)$,  the asymptotic quantities can be written as 
\begin{align}
\begin{split}
    M&=\frac{3\pi c \tilde{\kappa}^2}{1-c}\,,\qquad J_1=2\pi c \tilde{\kappa}^3\left(\frac{1+c}{1-c}\right)^{\frac{3}{2}}\,,\qquad J_2=0\,,\\
    \zeta&=0\,,\qquad \eta=\frac{3\pi \tilde{\kappa}^2(1-c+2c^2)}{2(1-c)}\,,
\end{split}
\end{align}
where we imposed the regularity condition (\ref{b-c-reg}).
For more details of the black ring solution, see for example \cite{Emparan:2001wn,Harmark:2004rm}.

\medskip

The rod structure is depicted in Fig.\,\ref{fig:er-ring-rod}, with three turning points,
\begin{align}\label{eq:ring_turning}
     w_1=-\frac{c}{2}\tilde\kappa^2\,,\qquad w_2=\frac{c}{2}\tilde\kappa^2,\qquad w_3=\frac{1}{2}\tilde\kappa^2
\end{align}
which consists of four rods: 
(i) the $\tilde \psi$-rotational axis: $I_1=\{(\rho,z)|\rho=0,-\infty<z<w_1\}$, (ii) the horizon cross section: $I_2=\{(\rho,z)|\rho=0,w_1<z<w_2\}$,  (iii)  the inner rotational axis of the ring: $I_3=\{(\rho,z)|\rho=0,w_2<z<w_3\}$, (iv) the $\tilde \phi$-rotational axis: $I_4=\{(\rho,z)|\rho=0,w_3<z<\infty\}$.
The rod vector on the finite interval $I_2$ takes the form $v_2=(1,\omega_{\tilde{\phi}},0)$ with the angular velocity of the horizon, $\omega_{\tilde{\phi}}=\frac{b-c}{(1-c)C\tilde{\kappa}}$. 
The rod vectors $v_1=(0,0,1)$ and $v_4=(0,1,0)$ on the semi-infinite rods indicate that $I_1$ and $I_4$ correspond to the fixed-point sets of the $U(1)$ isometries generated by $\partial_{\tilde{\psi}}$ and $\partial_{\tilde{\phi}}$, respectively. 
In addition, the finite rod $I_3$ with vector $v_3=(0,0,1)$ corresponds to the $\tilde\psi$-rotational axis inside the black ring.

\subsection{Coset space description}

Next, we derive the coset space description of the Emparan-Reall black ring solution (\ref{er-bring}).
As pointed out in Ref.~\cite{Chakrabarty:2014ora}, when one attempts to obtain the coset description corresponding to the metric~(\ref{er-bring}), certain components of the corresponding coset matrix diverge at spatial infinity.
This divergence prevents the straightforward application of the Riemann-Hilbert approach developed in~\cite{Katsimpouri:2012ky,Katsimpouri:2013wka}.
However, we will show that this technical difficulty can be circumvented by alternatively introducing the Euler angles $\phi$ and $\psi$ defined in Eq.~(\ref{new-angle}), in analogy with the treatment of the 5D Myers-Perry black hole.

In order to obtain the monodromy matrix, let us introduce the Weyl-Papapetrou coordinates $(\rho,z)$. To this end, we first define the associated Killing metric as
\begin{align}
    g_{\rm Killing}=(g_{ij})\,,\qquad i,j=t,\phi,\psi\,,
\end{align}
and its determinant is 
\begin{align}\label{det_g}
    {\rm det}\left(g_{{\rm Killing}}\right)=-\left(\frac{\tilde{\kappa}^2\sqrt{-G(u)G(v)}}{(u-v)^2}\right)^2\,.
\end{align}
From the relation (\ref{det_g}), we take the Weyl-Papapetrou coordinates $(\rho,z)$ as 
\begin{align}\label{weyl-er}
    \rho=\frac{\tilde{\kappa}^2\sqrt{-G(u)G(v)}}{(u-v)^2}\,,\qquad 
    z=\frac{\tilde{\kappa}^2(1-uv)(2+c(u+v))}{2(u-v)^2}\,.
\end{align}
For details on how $z$ is determined from this choice of $\rho$, see the appendix H in Ref.~\cite{Harmark:2004rm}.

The coset matrix $M_{\rm ER}(z,\rho)$, as in the examples discussed above, can be obtained via a generalized dimensional reduction to three dimensions with the metric
\begin{align}\label{3d-metric}
    ds_3^2=e^{2\nu}(d\rho^2+dz^2)+\rho^2d\phi^2\,.
\end{align}
The 16 scalar fields that parametrize the coset space are given by
\begin{align}\label{br-scalar}
\begin{split}
 e^{2U}&=\frac{\tilde{\kappa}^2}{2(u-v)^2}\sqrt{\frac{F(v)}{F(u)}}\left(G(u)-\frac{F(u)}{F(v)}G(v)\right)\,,\quad
    x^I=0\,,\quad y^I=\sqrt{\frac{F(v)}{F(u)}}\,,\\
   \zeta^0&=-\frac{\cC(1+v)}{2F(v)}\tilde{\kappa}\,,\quad
   \zeta^I=0\,,\quad   \tilde{\zeta}_{0}=\frac{\cC(1+u)}{2F(u)}\tilde{\kappa}\,,\qquad \tilde{\zeta}_{I}=0\,,\\
\sigma&=\tilde{\kappa}^2\biggl(-1+\frac{2c}{b}F(u)-\frac{2u+c(3u^2-1)+b(1+u^2+2c u^3)}{(u-v)F(u)}\\
&\qquad\qquad +\cC^2 \frac{(1-b)\left(4+2b(u+2v)+b^2(v-1+u(3v-1))\right)}{4b^2(1+b)F(u)F(v)}\biggr)\,,
\end{split}
\end{align}
and we obtain the corresponding coset matrix $M_{\rm ER}(z,\rho)$ from (\ref{iwasawa-rep}) and (\ref{m-def}).
We find that the coset matrix $M_{\rm ER}(z,\rho)$ approaches the constant matrix $Y_{\text{flat}}$ at the spacial infinity:
\begin{align}
    \lim_{r\to \infty}M_{\rm ER}(z,\rho)= Y_{\text{flat}}\,.
\end{align}
Furthermore, the conformal factor $e^{2\nu}$ can be read off from the three-dimensional metric (\ref{3d-metric}) obtained via the above dimensional reduction of (\ref{er-bring}). The explicit expression is
\begin{align}\label{er-conf}
    e^{2\nu}= \frac{4\left(u+v+b(1+uv)+c(-1+u^2+uv+v^2)+b c uv (u+v)\right)}{\left(2+c(1+u+v-uv)\right)\left(u+v+c(1+uv)\right)\left(2+c(-1+u+v+uv)\right)}\,.
\end{align}

\subsection{Monodromy matrix}

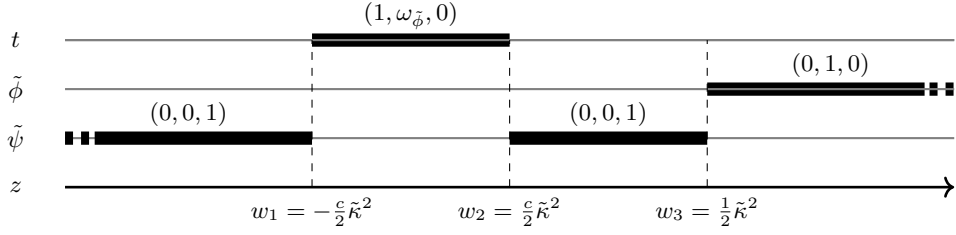
\begin{figure}
\begin{center}
\begin{tikzpicture}[scale=0.65]
\node[font=\small ] at (-10,3) {$t$};
\node[font=\small ] at (-10,2) {$\tilde{\phi}$};
\node[font=\small ] at (-10,1) {$\tilde{\psi}$};
\node[font=\small ] at (-10,0) {$z$};
\node[font=\small ] at (-6.5,1.5) {$(0,0,1)$};
\node[font=\small ] at (2,1.5) {$(0,0,1)$};
\node[font=\small ] at (6.5,2.5) {$(0,1,0)$};
\node[font=\small ] at (-2,3.5) {$(1,\omega_{\tilde{\phi}},0)$};
\node[font=\small ] at (-4,-0.5) {$w_1=-\frac{c}{2}\tilde{\kappa}^2$};
\node[font=\small ] at (0,-0.5) {$w_2=\frac{c}{2}\tilde{\kappa}^2$};
\node[font=\small ] at (4,-0.5) {$w_3=\frac{1}{2}\tilde{\kappa}^2$};
\draw[gray,line width = 0.8] (-9,3) -- (9,3);
\draw[black,line width = 5] (-4,3) -- (0,3);
\draw[gray,line width = 0.8] (-9,3) -- (9,3);
\draw[gray,line width = 0.8] (-9,1) -- (9,1);
\draw[black,line width = 5] (4,2) -- (8.4,2);
\draw[black,line width = 5,dashed ] (8.5,2) -- (9,2);
\draw[gray,line width = 0.8] (-9,2) -- (9,2);
\draw[black,line width = 5,dashed ] (-9,1) -- (-8.5,1);
\draw[black,line width = 5] (-8.4,1) -- (-4,1);
\draw[black,line width = 5] (0,1) -- (4,1);
\draw[black,dashed ] (-4,0) -- (-4,3);
\draw[black,dashed ] (0,0) -- (0,3);
\draw[black,dashed ] (4,0) -- (4,3);
\draw[->,black,line width = 1] (-9,0) -- (9,0);
\end{tikzpicture}
\caption{The rod diagram for the 5D Emparan-Reall black ring. The corner points $w_i$ satisfy $w_1<w_2<w_3$.
}
\label{fig:er-ring-rod}
\end{center}
\end{figure}

We now derive the monodromy matrix $\cM_{\rm ER}(w)$ corresponding to the Emparan-
Reall black ring. Following Ref.~\cite{Harmark:2004rm}, we transform the $C$-metric coordinates $(u,v)$ into the Weyl-
Papapetrou coordinates $(\rho,z)$, where the relations are given by
\begin{align}
    u=-\frac{R_1-R_2+2R_3-\tilde{\kappa}^2 }{R_1+R_2-c\tilde{\kappa}^2}\,,\qquad
    v=-\frac{R_1-R_2+2R_3+\tilde{\kappa}^2 }{R_1+R_2+c\tilde{\kappa}^2}\,,
\end{align}
with
\begin{align}
    R_i=\sqrt{\rho^2+(z-w_i)^2}\,,
\end{align}
where $w_i\ (i=1,2,3)$ are defined by Eq.~(\ref{eq:ring_turning}).
By using the relation (\ref{sub-rule}), we can find that the monodromy matrix $\cM_{\rm ER}(w)$ takes the form
\begin{align}\label{br-mono}
    \cM_{\rm ER}(w)=Y_{\rm flat}+\sum_{i=1}^{3}\frac{A_i}{w-w_i}\,,
\end{align}
where it should be noted that the locations of the poles $w_i$ coincide with the turning points of the rods in Fig.~\ref{fig:er-ring-rod}.
The explicit expressions of the residue matrices $A_j$ are given by
\begin{align}
\begin{split}\label{er-a}
     A_1&=\frac{1-c}{(1-b)(1+c)}\tilde{\kappa}^2
    \begin{pmatrix}
       -\frac{b(1+b)(1-c)}{1-b}&0&0&-\frac{\cC}{\tilde{\kappa}}&0&0&\cC\frac{1-c}{2(1-b)}\tilde{\kappa}&0\\
        0&0&0&0&0&0&0&0\\
      0&0&-\frac{2c}{\tilde{\kappa}^2}\frac{1-b}{1-c}&0&0&0&0&-c\\
      \frac{\cC}{\tilde{\kappa}}&0&0&\frac{\cC^2}{\tilde{\kappa}^2}\frac{1-b}{b(1+b)(1-c)}&0&0&-\cC^2\frac{1}{2b(1+b)}&0\\
        0&0&0&0&0&0&0&0\\
        0&0&0&0&0&0&0&0\\
       \cC\frac{1-c}{2(1-b)}\tilde{\kappa}&0&0&\cC^2\frac{1}{2b(1+b)}&0&0&-\cC^2\frac{1-c}{4b(1-b^2)}\tilde{\kappa}^2&0\\
       0&0&c&0&0&0&0&\frac{c(1-c)}{2(1-b)}\tilde{\kappa}^2\\
    \end{pmatrix}\,,\\
     A_2&=\frac{1-c}{1-b}\tilde{\kappa}^2\begin{pmatrix}
       0&0&0&0&0&0&0&0\\
        0&0&0&0&0&0&0&0\\
      0&0&0&0&0&0&0&0\\
      0&0&0&0&0&0&0&0\\
      0&0&0&0&b&0&0&-\frac{1}{2}\cC \tilde{\kappa}\\
        0&0&0&0&0&0&0&0\\
       0&0&0&0&0&0&\frac{1}{2}(1-b)c \tilde{\kappa}^4&0\\
       0&0&0&0&\frac{1}{2}\cC \tilde{\kappa}&0&0&-\frac{1}{4b}\cC^2 \tilde{\kappa}^2\\
    \end{pmatrix}\,,\\
    A_3&=\frac{1-c}{(1-b)(1+c)}\tilde{\kappa}^2
    \begin{pmatrix}
       \frac{\cC^2}{1+b}&0&0&\frac{\cC}{\tilde{\kappa}}&0&0&-\cC\frac{b-2c+bc}{2(1-b)}\tilde{\kappa}&0\\
        0&0&0&0&0&0&0&0\\
      0&0&-\frac{1-b}{\tilde{\kappa}^2}&0&0&0&0&-\frac{b-2c+ bc}{2}\\
      -\frac{\cC}{\tilde{\kappa}}&0&0&-\frac{1+b}{\tilde{\kappa}^2}&0&0&\frac{(1+b)(b-2c+bc)}{2(1-b)}&0\\
        0&0&0&0&0&0&0&0\\
        0&0&0&0&0&0&0&0\\
       -\cC\frac{b-2c+bc}{2(1-b)}\tilde{\kappa}&0&0&-\frac{(1+b)(b-2c+bc)}{2(1-b)}&0&0&\frac{(1+b)(b-2c+bc)^2}{4(1-b)^2}\tilde{\kappa}^2&0\\
       0&0&\frac{b-2c+bc}{2}&0&0&0&0&\frac{(b-2c+bc)^2}{4(1-b)}\tilde{\kappa}^2\\
    \end{pmatrix}\,.
\end{split}
\end{align}
The residue matrices $A_j$ are rank 2, and hence the Riemann-Hilbert approach developed in \cite{Katsimpouri:2012ky,Katsimpouri:2013wka} can be applied to this solution.

\subsubsection*{Charge matrix}

As in the spherical case, we compute the charge matrix for the Emparan-Reall black ring.
While the charge matrix is useful in the classification of extremal limits of black holes with spherical horizon topology, its relevance for the non-spherical cases is not clear.
Here, we write down the explicit expression of the charge matrix and discuss how its algebraic relations relate to the underlying geometric structure.

The charge matrix $Q$ can be read off the large-$w$ behavior of the monodromy matrix (\ref{br-mono}) and is given by
\begin{align}
     Q=Y_{\rm flat}^{-1}\left(\sum_{j=1}^{3}A_j\right)\,.
\end{align}
By substituting (\ref{er-a}) into this, the explicit expression is 
\begin{align}\label{ring-cherge}
    Q&=-\frac{b(1-c)}{2(1-b)}\tilde{\kappa}^2 \sum_{j=1}^{3}H_j
    +\frac{(1-c)(-b^2(1-c)+4c-2b(1+c))}{4(1-b)^2}\tilde{\kappa}^4E_0\no\\
    &\quad+F_0+\frac{\cC}{2}\frac{1-c}{1-b}\tilde{\kappa}^3(E_{p_0}+E_{q^0})\,.
\end{align}
We find that the expression matches the universal form (\ref{Q-gen}) with the asymptotic quantities
\begin{align}
\begin{split}
    M&=\frac{3\pi b(1-c) \tilde{\kappa}^2}{2(1-b)}\,,\qquad J_1=\pi \cC \tilde{\kappa}^3\frac{1-c}{1-b}\,,\qquad J_2=0\,.
\end{split}
\end{align}
The charge matrix (\ref{ring-cherge}) does not satisfy the cubic equation (\ref{mp-q-rel}) associated with the Myers-Perry black hole solution, but satisfy a slightly modified equation with additional terms involving the Cartan generators
\begin{align}
\begin{split}\label{ring-cubic}
    &Q^3-\frac{1}{4}\Tr(Q^2)Q+\frac{bc (1-c)^2}{2(1-b)}\tilde{\kappa}^{6} \left(H_1-\frac{1}{2}H_2-\frac{1}{2}H_3\right)=0\,,\\
    &\Tr(Q^2)=2\frac{1-b}{1-c}(2c-b(1-c))\tilde{\kappa}^4\,.
\end{split}
\end{align}
When we impose the regularity condition (\ref{b-c-reg}) on $Q$ and denote it by $\mathcal{Q}$, the equation (\ref{ring-cubic}) reduces to
\begin{align}
\begin{split}
    &\mathcal{Q}^3-\frac{1}{4}\Tr(\mathcal{Q}^2)\mathcal{Q}+c^2 \tilde{\kappa}^6\left(H_1-\frac{1}{2}H_2-\frac{1}{2}H_3\right)=0\,,\\
    &\Tr(\mathcal{Q}^2)=4c^2\frac{1+c}{1-c}\tilde{\kappa}^4\,.
\end{split}
\end{align}
We find that the condition $\Tr(\mathcal{Q}^2)=0$ is satisfied only at a point $c=0$ because $0<c<1$ and the associated monodromy matrix is
\begin{align}
    \lim_{c\to0}\left(\cM_{\rm ER}(w)\lvert_{(\ref{b-c-reg})}\right)=Y_{\rm flat}\left(1+\frac{F_0}{w-w_3}\right)\,.
\end{align}
This is equivalent to the monodromy matrix (\ref{one-mono2}) corresponding to five-dimensional Minkowski spacetime. Indeed, the Emparan-Reall black ring (\ref{er-bring}) with the balanced condition (\ref{b-c-reg}) and $c=0$ reduces to the flat spacetime.
This result is consistent with the nonexistence of the extremal limit for the Emparan-Reall black ring.

\subsection{Factorization of monodromy matrix}

We explicitly perform the factorization of the monodromy matrix (\ref{br-mono}).
To this end, we again express the residue matrices $A_j$ in terms of the eight-component vectors $a_j$ and $b_j$ using the eigenvectors with the non-zero eigenvalues; the explicit expressions are shown as
\begin{align}
\begin{split}
    a_1^T&=\left(0,0,-\frac{2(1-b)}{\tilde{\kappa}^2(1-c)},0,0,0,0,1\right)\,,\\
    b_1^T&=\left(1,0,0,-\cC\frac{1-b}{b(1+b)(1-c)\tilde{\kappa}},0,0,-\cC\frac{\tilde{\kappa}}{2b(1+b)},0\right)\eta\,,\\
    a_2^T&=\left(0,0,0,0,1,0,0,\cC\frac{\tilde{\kappa}}{2b}\right)\,,\\
    b_2^T&=\left(0,0,0,0,0,0,1,0\right)\eta\,,\\
    a_3^T&=\left(0,0,-\frac{2(1-b)}{\tilde{\kappa}^2(b-2c+b c)},0,0,0,0,1\right)\,,\\
    b_3^T&=\left(-\frac{4b(b-c)}{\tilde{\kappa}(b-2c+b c)\cC},0,0,\frac{2(1-b)}{\tilde{\kappa}^2(b-2c+b c)},0,0,\cC,0\right)\eta\,,
\end{split}
\end{align}
and the constants $\alpha_j,\beta_j$ are given by
\begin{align}
\begin{split}
    \alpha_1&=\frac{(1-c)^2c}{2(1-b)^2(1+c)}\tilde{\kappa}^4\,,\qquad 
    \beta_1=-\frac{b(1+b)(1-c)^2}{(1+c)(1-b)^2}\tilde{\kappa}^2\,,\\
    \alpha_2&=-\frac{b(1-c)}{1-b}\tilde{\kappa}^2\,,\qquad \beta_2=\frac{1}{2}c(1-c)\tilde{\kappa}^4\,,\\
    \alpha_3&=\frac{(1-c)(b-2c+bc)^2}{4(1-b)^2(1+c)}\tilde{\kappa}^4\,,\qquad 
    \beta_3=\frac{(1-c)(1+b)(b-2c+bc)^2}{4(1-b)^3(1+c)}\tilde{\kappa}^4\,.
\end{split}
\end{align}
In this choice of the vectors, the $3\times 3$ matrices $\Gamma^{(0)}$ and $\Gamma^{(a)},\Gamma^{(b)}$ become
\begin{align}
    \Gamma^{(0)}&=
    \begin{pmatrix}
        \cC\frac{2(1-b)^2}{b(1+b)(1-c)^2\tilde{\kappa}^3}\frac{1}{\la_1\nu_1}&-\frac{2(1-b)}{(1-c)\tilde{\kappa}^2}\frac{1}{\la_{1,2}}&\cC \frac{2(1-b)^2(1+c)}{(1-c)(b-2c+bc)\tilde{\kappa}^2}\frac{1}{\la_{1,3}}\\
        -\frac{(1-b)c}{(1-c)b}\frac{1}{\la_{1,2}}&0&\cC\frac{b^2-2b+1}{b(1+b)(b-2c+bc)\tilde{\kappa}}\frac{1}{\la_{2,3}}\\
        -\cC \frac{(b^2-2b+1)(1+c)}{b(1+b)(1-c)(b-2c+bc)\tilde{\kappa}}\frac{1}{\la_{1,3}}&\frac{2(1-b)}{(b-2c+bc)\tilde{\kappa}^2}\frac{1}{\la_{2,3}}&-\frac{4(1-b)^2}{(b-2c+bc)^2\tilde{\kappa}^4}\frac{1}{\la_3\nu_3 }
    \end{pmatrix}\,,\label{er-gamma}\\
    \Gamma^{(a)}&=\Gamma^{(b)}=0_{3\times 3}\,.
\end{align}
By using the relations (\ref{xp-ex}) and (\ref{c-def}), we can construct the matrix $X_+$ and show that the monodromy matrix $\cM_{\rm ER}(w)$ can be factorized into the form
\begin{align}\label{mon-fac-er}
    \cM_{\rm ER}(w(\la,z,\rho))=X_-(\la,z,\rho)M_{\rm ER}(z,\rho)X_+(\la,z,\rho)\,.
\end{align}
therefore, the monodromy matrix (\ref{br-mono}) precisely describes the Emparan-Reall black ring solution (\ref{er-bring}).

\subsubsection*{Conformal factor}

Next, we evaluate the conformal factor $e^{2\nu}$.
Since $\Gamma^{(a)}=\Gamma^{(b)}=0_{3\times 3}$, we can again employ the formula (\ref{conf}).
From the expression (\ref{er-gamma}) of $\Gamma^{(0)}$, the right-hand side can be computed as
\begin{align}\label{conf-er-com}
   &k_{\rm BM}\prod_{j=1}^{3}(\la_j\nu_j)\,{\rm det}(\Gamma^{(0)})
    =k_{\rm BM}\left(\frac{8(1-b)^4(1+c)^2}{bc(1+b)(1-c)^2(b-2c+bc)^2\tilde{\kappa}^{10}}\right)\no\\
    &\times \frac{4\left(u+v+b(1+uv)+c(-1+u^2+uv+v^2)+b c uv (u+v)\right)}{\left(2+c(1+u+v-uv)\right)\left(u+v+c(1+uv)\right)\left(2+c(-1+u+v+uv)\right)}\,.
\end{align}
Fixing the overall constant $k_{\rm BM}$ as
\begin{align}
    &k_{\rm BM}^{-1}=\frac{8(1-b)^4(1+c)^2}{bc(1+b)(1-c)^2(b-2c+bc)^2\tilde{\kappa}^{10}}
\end{align}
precisely reproduces the conformal factor (\ref{er-conf}) for the 5D rotating black ring solution. 

\subsubsection*{Static limit}

Finally, we give a comment on the static limit of the rotating black ring (\ref{er-bring}).
This can be realized by taking a limit $b\to c$ i.e. $\cC\to 0$. In this limit, both the factorization (\ref{mon-fac-er}) and the conformal factor (\ref{conf-er-com}) still hold.
Thus, the monodromy matrix (\ref{br-mono}) can be regarded as encoding the physical information of the black ring with a single angular momentum.

\subsection{From 5D rotating black ring to 5D Myers-Perry}

It is known that the 5D Myers-Perry black hole can be obtained as a scaling limit of the 5D singly rotating black ring \cite{Emparan:2004wy}. Here, we investigate how this relation is realized at the level of the monodromy matrix.

\subsubsection{Scaling limit of 5D rotating black ring metric}

To this end, we first review how the scaling limit of the 5D rotating black ring solution (\ref{er-bring}) reduces to the 5D Myers-Perry black hole solution (\ref{mr-metric}).
This can be accomplished by considering a limit \cite{Emparan:2004wy}
\begin{align}\label{lim-mp-br}
   b\,, c \to 1\,,\qquad  \tilde{\kappa}\to0
\end{align}
with the following ratios fixed:
\begin{align}
    r_0^2 =\frac{4\tilde{\kappa}^2}{1-c}\,,\qquad a^2=4\tilde{\kappa}^2\frac{b-c}{(1-c)^2}\,.
\end{align}
This scaling limit can be implemented as a small $\epsilon$ limit by redefining the parameters $b\,, c\,, \tilde{\kappa}$ as
\begin{align}\label{redef}
    c=1-\epsilon\,,\qquad b=1-\epsilon \cos^2\la\,,\qquad \tilde{\kappa}=\frac{\sqrt{\alpha}}{\cos\la}\sqrt{\epsilon}\,.
\end{align}
In this limit, we don't require the condition for the absence of a conical singularity on the finite rod in $z\in [w_2,w_3]$, and only impose the following regularity conditions for the periodicity of the angle variables on the intervals $(-\infty,w_1]$ and $[w_3,\infty)$:
\begin{align}
    \Delta\tilde{\phi}=\Delta\tilde{\psi} =2\pi \frac{\sqrt{1-b}}{1-c}\,.
\end{align}
By taking care of the regularity conditions, we rescale them as
\begin{align}\label{angle-def}
    (\tilde{\psi},\tilde{\phi})= \sqrt{\frac{r_0^2-a^2}{4\tilde{\kappa}^2}}(\tilde{\psi}',\tilde{\phi}')=\frac{\cos\la}{\sqrt{\epsilon}}(\tilde{\psi}',\tilde{\phi}')\,,
\end{align}
so that the new angular variables $\tilde{\psi}',\tilde{\phi}'$ have the period $2\pi$.
We also make a change of the coordinates $(u,v)$ in the rotating black ring solution (\ref{er-bring}) to the spherical coordinates $(r,\theta)$ as follows \cite{Emparan:2004wy}
\begin{align}
\begin{split}
    u&=-1+2\left(1-\frac{a^2}{r_0^2}\right)\frac{2\tilde{\kappa}^2\cos^2\theta}{r^2-(r_0^2-a^2)\cos^2\theta}\,,\\
    v&=-1-2\left(1-\frac{a^2}{r_0^2}\right)\frac{2\tilde{\kappa}^2\sin^2\theta}{r^2-(r_0^2-a^2)\cos^2\theta}\,.
\end{split}
\end{align}
By taking the limit $\epsilon\to 0$, we can see that the 5D rotating black ring solution (\ref{er-bring}) becomes to the 5D Myers-Perry black hole solution (\ref{mr-metric}).

\subsubsection{Limit of monodromy matrix}

Let us now examine how the degenerate limit discussed above is realized in the monodromy matrix. We find that the redefinition (\ref{angle-def}) of the angle variables results in a replacement of $\tilde{\kappa}$ with $\alpha^{1/2}$ in $M_{\rm ER}(z,\rho)$ :
\begin{align}
   M_{\rm ER}'(z,\rho)= M_{\rm ER}(z,\rho)\lvert_{\tilde{\kappa} \to \alpha^{1/2}}\,.
\end{align}
As a consequence, this substitution also modifies the positions of the poles in $\cM_{\rm ER}(w)$:
\begin{align}
\begin{split}\label{pole-br-re}
      w_1&=-\frac{1}{2}c\tilde{\kappa}^2\,,\qquad w_2=\frac{1}{2}c\tilde{\kappa}^2\,,\qquad w_3=\frac{1}{2}\tilde{\kappa}^2\,,\\
     \to w_1'&=-\frac{1}{2}c\alpha\,,\qquad w_2'=\frac{1}{2}c\alpha\,,\qquad w_3'=\frac{1}{2}\alpha\,.
\end{split}
\end{align}
Taking a limit $\epsilon\to0$ with the redefinition (\ref{redef}), the rescaled simple poles $w'_i$ become two simple poles 
\begin{align}
    \tilde{w}_1=\lim_{\epsilon\to 0}w_1'=-\frac{1}{2}\alpha\,,\qquad \tilde{w}_2=\lim_{\epsilon\to0}w_2'=\lim_{\epsilon\to0}w_3'=\frac{1}{2}\alpha\,.
\end{align}
This degeneration of simple poles is consistent with the transition of the rod structure (Fig.\,\ref{fig:er-ring-rod}) of the black ring to that (Fig.\,\ref{mp-rod}) of the Myers-Perry black hole. We can also find that the modified reside matrices are reduced to the residue matrices (\ref{mp-A}) for the Myers-Perry black hole as follows:
\begin{align}
    \tilde{A}_1&=\lim_{\epsilon\to 0} A_1\lvert_{\tilde{\kappa} \to \alpha^{1/2}}\,,\\
    \tilde{A}_2&=\lim_{\epsilon\to 0}\left(A_2+A_3\right)\lvert_{\tilde{\kappa} \to \alpha^{1/2}}\,,   
\end{align}
where we have relabeled the residue matrices (\ref{mp-A}) as $\tilde{A}_i$.
Thus, the monodromy matrix $\cM_{\rm ER}(w)$ reduces to $\cM_{\rm MP}(w)$ in the small $\epsilon$ limit,
\begin{align}
    \cM_{\rm MP}(w)=\lim_{\epsilon\to 0}\cM_{\rm ER}(w)\lvert_{\tilde{\kappa} \to \alpha^{1/2}}\,.
\end{align}
Finally, we consider the scaling limit of the conformal factor (\ref{er-conf}). By taking about the rescale (\ref{angle-def}) of $\tilde{\phi}$, we take a limit with (\ref{redef})
\begin{align}
   e^{2\nu_{\rm MP}}=\lim_{\epsilon\to0} \frac{\epsilon}{\cos^2\la} e^{2\nu_{\rm ER}}\,.
\end{align}

\newpage

\section{Rotating black lens}\label{sec:ct}

Several attempts were made using the inverse scattering method to construct asymptotically flat solutions to the five-dimensional vacuum Einstein equations, but all such efforts proved unsuccessful~\cite{Evslin:2008gx,Chen:2008fa,Tomizawa:2019acu}.
The central difficulty in realizing a regular black lens lies in the fact that the candidate solutions invariably contain naked singularities.
With the aim of paving the way toward a future construction of regular vacuum black lens solutions, we here derive the monodromy matrix corresponding to the five-dimensional black lens with a single angular momentum, originally obtained by Chen and Teo~\cite{Chen:2008fa}, even though their solution itself is singular.

\subsection{Solution}

The metric for the black lens solution with a single angular momentum can be written in the $C$-metric form, as originally constructed by Chen and Teo
\begin{align}\label{rrrlens-f}
    ds_5^2&=-\frac{H(v,u)}{H(u,v)}\left(dt-\Omega_{\tilde{\phi}}d\tilde{\phi}-\Omega_{\tilde{\psi}}d\tilde{\psi}\right)^2
    +\frac{\tilde{\kappa}^2H(u,v)}{2(1-a^2)(1-b)^3(u-v)^2}
    \left(\frac{du^2}{G(u)}-\frac{dv^2}{G(v)}\right)\no\\
   &\qquad +\frac{F(v,u)}{H(v,u)}d\tilde{\psi}^2
   -\frac{F(u,v)}{H(v,u)}d\tilde{\phi}^2+2\frac{J(u,v)}{H(v,u)}d\tilde{\phi}d\tilde{\psi}\,.
\end{align}
where $\Omega_{\tilde{\phi}}$ and $\Omega_{\tilde{\psi}}$ are given by
\begin{align}
    \Omega_{\tilde{\phi}}(u,v)&=\frac{\Omega_0(1+v)}{H(v,u)}
    \Bigl(2 (1-c) \left(1-b-a^2(1+b u)\right)^2\no\\
    &\qquad\qquad\qquad-a^2 \left(1-a^2\right) b (1-b) (1-u) (1+v) (1+c u)\Bigr)\,,\\
    \Omega_{\tilde{\psi}}(u,v)&=\frac{\Omega_0a(1+u)^2 (1+v)}{H(v,u)} \left(a^4 (b+1) (b-c)+a^2 (1-b) (b c-b+2 c)-(1-b)^2 c\right)\,,\\
    \Omega_0&=2\tilde{\kappa}(1-c)\sqrt{\frac{2b(1+b)(b-c)}{(1-a^2)(1-b)}}\,,
\end{align}
and the functions $G\,,H\,, F\,,$ and $J$ are defined as
\begin{align}
    G(u)&=(1-u^2)(1+cu)\,,\\
    H(u,v)&=a^2(b-c)(1+u)(1+v)
    \Bigl(-2 b (1-b) (1-c) (1-u)\no\\
    &\quad+(1+b) (1+v) \left( c\left(1-a^2\right) (1-b)(1+u)
    +2 a^2 b (1-c)\right)\Bigr)\no\\
    &\quad+4 (1-b) (1-c) (1+bu) \left((1-b) (1-c)-a^2 ((1+bu) (1+cv)+ (b-c)(1+v))\right)\,,\\
    F(u,v)&=\frac{2 \tilde{\kappa}^2}{\left(1-a^2\right) (u-v)^2}\biggl[4(1-c)^2 (1+b u) \left(1-b-a^2(1+b u)\right)^2G(v)\no\\
    &\quad-a^2G(u)(1+v)^2 \biggl((1-c)^2 \left(1-b-a^2(b+1)\right)^2 (1+b v)
    -\left(1-a^2\right) \left(1-b^2\right) \no\\
    &\quad\times(1+c v) \left( (b-c)\left(1-a^2\right)(1+v) 
    +(1-c) \left(1-3b-a^2(b+1)\right)\right)\biggr)\biggr]\,,\\
    J(u,v)&=\frac{4\tilde{\kappa}^2 a (1-c)  (1+u) (1+v)}{\left(1-a^2\right) (u-v)}\left(1-b-a^2(b+1)\right) \left((1-b) c+a^2 (b-c)\right)\no\\
    &\qquad\times \Bigl[(1+b u) (1+c v)+(1+c u)(1+b v)+(b-c) (1-u v)\Bigr]\,,
\end{align}
where the $C$-metric coordinates $u$ and $v$ has the ranges:
\begin{align}
    -1\leq u \leq 1\,,\qquad -\frac{1}{c}< v \leq -1\,,
\end{align}
and the parameters $a,b$, and $c$  are restricted to
\begin{align}
    -1<a<1\,,\qquad 0<c\leq  b<1\,.
\end{align}

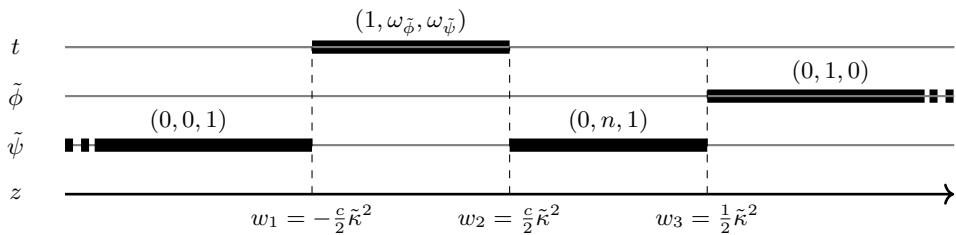
\begin{figure}
\begin{center}
\begin{tikzpicture}[scale=0.65]
\node[font=\small ] at (-10,3) {$t$};
\node[font=\small ] at (-10,2) {$\tilde{\phi}$};
\node[font=\small ] at (-10,1) {$\tilde{\psi}$};
\node[font=\small ] at (-10,0) {$z$};
\node[font=\small ] at (-6.5,1.5) {$(0,0,1)$};
\node[font=\small ] at (2,1.5) {$(0,n,1)$};
\node[font=\small ] at (6.5,2.5) {$(0,1,0)$};
\node[font=\small ] at (-2,3.5) {$(1,\omega_{\tilde{\phi}},\omega_{\tilde{\psi}})$};
\node[font=\small ] at (-4,-0.5) {$w_1=-\frac{c}{2}\tilde{\kappa}^2$};
\node[font=\small ] at (0,-0.5) {$w_2=\frac{c}{2}\tilde{\kappa}^2$};
\node[font=\small ] at (4,-0.5) {$w_3=\frac{1}{2}\tilde{\kappa}^2$};
\draw[gray,line width = 0.8] (-9,3) -- (9,3);
\draw[black,line width = 5] (-4,3) -- (0,3);
\draw[gray,line width = 0.8] (-9,3) -- (9,3);
\draw[gray,line width = 0.8] (-9,1) -- (9,1);
\draw[black,line width = 5] (4,2) -- (8.4,2);
\draw[black,line width = 5,dashed ] (8.5,2) -- (9,2);
\draw[gray,line width = 0.8] (-9,2) -- (9,2);
\draw[black,line width = 5,dashed ] (-9,1) -- (-8.5,1);
\draw[black,line width = 5] (-8.4,1) -- (-4,1);
\draw[black,line width = 5] (0,1) -- (4,1);
\draw[black,dashed ] (-4,0) -- (-4,3);
\draw[black,dashed ] (0,0) -- (0,3);
\draw[black,dashed ] (4,0) -- (4,3);
\draw[->,black,line width = 1] (-9,0) -- (9,0);
\end{tikzpicture}
\caption{Rod diagram for 5D rotating black lens. The corner points satisfy $w_1<w_2<w_3$. }
\label{fig:rlens-rod}
\end{center}
\end{figure}

The corresponding rod structure, as shown in Fig.\ref{fig:rlens-rod} is divided by three turning points:
\begin{align}\label{eq:lens_turning}
     w_1=-\frac{c}{2}\tilde\kappa^2\,,\qquad w_2=\frac{c}{2}\tilde\kappa^2,\qquad w_3=\frac{1}{2}\tilde\kappa^2
\end{align}
 into four parts: (i) the $\tilde \psi$-rotational axis: $I_1=\{(\rho,z)|\rho=0,-\infty<z<w_1\}$, (ii) the horizon cross section: $I_2=\{(\rho,z)|\rho=0,w_1<z<w_2\}$,  (iii)  the inner rotational axis: $I_3=\{(\rho,z)|\rho=0,w_2<z<w_3\}$, (iv) the $\tilde \phi$-rotational axis: $I_4=\{(\rho,z)|\rho=0,w_3<z<\infty\}$.
The rod vector on the finite interval $I_2$ takes the form $v_2=(1,\omega_{\tilde{\phi}},\omega_{\tilde{\psi}})$, which are the non-zero angular velocities of the horizon
\begin{align}
    \omega_{\tilde{\phi}}&=\frac{1}{\tilde{\kappa}}\sqrt{\frac{(1-b)(b-c)}{2(1-a^2)b(1+b)}}\frac{1}{1-c}\,,\qquad
    \omega_{\tilde{\psi}}=\frac{1}{2\tilde{\kappa}}\sqrt{\frac{(1-b)(b-c)}{2(1-a^2)b(1+b)}}\frac{a(1-b-a^2(1+b))}{(1-b)c+a^2(b-c)}\,.
\end{align}
The rod vectors $v_1=(0,0,1)$ and $v_4=(0,1,0)$ on the semi-infinite rods indicate that $I_1$ and $I_4$ correspond to the fixed-point sets of the $U(1)$ isometries generated by $\partial_{\tilde{\psi}}$ and $\partial_{\tilde{\phi}}$, respectively. 
In addition, the finite rod $I_3$ with vector $v_3=(0,n,1)$ corresponds to the fixed-point set of the $U(1)$ isometry generated by $n\partial_{\tilde{\phi}}+\partial_{\tilde{\psi}}$, where the integer $n$ is given by
\begin{align}
    n=\frac{2a((1-b)c+a^2(b-c))}{(1-b-a^2(1+b))(1-c)}\,,\qquad |n|=2,3,\ldots\,.
\end{align}
The two rod vectors $v_1$ and $v_3$ satisfy $|{\rm det}(v_1,v_3)|=n$, which implies that the topology of the horizon cross section is the lens space $L(n;1)$.

\medskip
Finally, the ADM mass and angular momenta are expressed as
\begin{align}
\begin{split}\label{asym-lens}
    M&=\frac{3\pi \tilde{\kappa}^2 b(1-c)}{2(1-b)}\,,\qquad J_1=\frac{\pi  \tilde{\kappa}^3\sqrt{2(1-a^2)b(1+b)(b-c)}(1-c)}{(1-b)^{\frac{3}{2}}}\,,\qquad J_2=0\,.
\end{split}
\end{align}

\subsection{Coset space description}

Let us compute the coset matrix $M_{\rm CT}(\rho,z)$ for the rotating black lens solution with a single angular momentum, given in Ref.~(\ref{rrrlens-f}).
As in the previous two examples, we employ the Euler angles~(\ref{new-angle}) instead of the original angular coordinates $(\tilde{\phi},\tilde{\psi})$.
We also adopt the same Weyl-
Papapetrou coordinates $(\rho,z)$ defined in (\ref{weyl-er}) as in the black ring case.
By performing a dimensional reduction to three dimensions and dualizing the one-form fields into scalar fields, we obtain the 16 scalar fields expressed as
\begin{align}
\begin{split}
 e^{2U}&=\frac{F(v,u)-F(u,v)-2J(u,v)}{4\sqrt{H(u,v)H(v,u)}}\,,\\
    x^I&=0\,,\quad y^I=\sqrt{\frac{H(v,u)}{H(u,v)}}\,,\\
   \zeta^{0}&=\frac{\Omega_{\tilde{\psi}}(u,v)-\Omega_{\tilde{\phi}}(u,v)}{2} \,,\qquad
   \tilde{\zeta}_{0}=-\frac{\Omega_{\tilde{\psi}}(v,u)-\Omega_{\tilde{\phi}}(v,u)}{2}\,,\\
   \zeta^{I}&=0\,,\qquad \tilde{\zeta}_{I}=0\,,\\
\sigma&=\frac{(1-a)\tilde{\kappa}^2}{4(1+a)(u-v)H(u,v)H(v,u)}\sigma_0(u,v)\,.
\end{split}\label{lens-scalar}
\end{align}
Here, $\sigma_0(u,v)$ is a symmetric polynomial of degree five in $u$ and $v$. We cannot find a compact expression of $\sigma_0(u,v)$, and its explicit expression is presented in appendix \ref{sec:twist}.
We find that the coset matrix $M_{\rm CT}(z,\rho)$ with (\ref{lens-scalar}) approaches the constant matrix $Y_{\text{flat}}$ at the spacial infinity:
\begin{align}
    \lim_{r\to \infty}M_{\rm CT}(z,\rho)= Y_{\text{flat}}\,.
\end{align}
Furthermore, the conformal factor $e^{2\nu}$ is given by
\begin{align}\label{conf-lens}
    e^{2\nu}
    &=\frac{1}{2(1-a^2)(1-b)^3\tilde{\kappa}^2}\frac{(u-v)\left(F(v,u)-F(u,v)-2J(u,v)\right)}{(u+v+c(1+uv))(2+c(1+u+v-uv))(2+c(-1+u+v+uv))}\,.
\end{align}

\subsection{Monodromy matrix}

Now let us evaluate the monodromy matrix corresponding to the 5D rotating black lens solution (\ref{rrrlens-f}).
By using the relation (\ref{sub-rule}), we can find the monodromy matrix $\cM_{\rm CT}(w)$ for the black lens with one angular momentum solution given by
\begin{align}\label{ct-mono}
    \cM_{\rm CT}(w)=Y_{\rm flat}+\sum_{i=1}^{3}\frac{A_i}{w-w_i}\,,
\end{align}
where the positions of the poles are
\begin{align}
     w_1=-\frac{1}{2}c\tilde{\kappa}^2\,,\qquad w_2=\frac{1}{2}c\tilde{\kappa}^2\,,\qquad w_3=\frac{1}{2}\tilde{\kappa}^2\,.
\end{align}
The explicit expressions of the residue matrices $A_j$ are given by
\begin{scriptsize}
\begin{align}
\begin{split}
     A_1&=
  \left(
\begin{array}{cccccccc}
 -\frac{b (b+1) (1-c)^2 S_1 \tilde{\kappa}^2}{(1-b)^3 c (c+1)} & 0 & 0 & -\frac{ \Omega_0S_1}{2 (1-b)^2 c (c+1)} & 0 & 0 & \frac{\Omega_0 S_0S_1 (1-a) (1-c) \tilde{\kappa}^2}{4 (a+1) (1-b)^3 c (c+1)}  & 0 \\
 0 & 0 & 0 & 0 & 0 & 0 & 0 & 0 \\
 0 & 0 & -\frac{2S_1}{\left(1-a^2\right) (1-b) (c+1)} & 0 & 0 & 0 & 0 & -\frac{(1-c) S_0S_1 \tilde{\kappa}^2 }{(a+1)^2 (1-b)^2 (c+1)} \\
 \frac{ \Omega_0S_1}{2 (1-b)^2 c (c+1)} & 0 & 0 & \frac{2 (b-c) S_1}{\left(1-a^2\right) (1-b)^2 c (c+1)} & 0 & 0 & -\frac{(b-c)(1-c) S_0S_1\tilde{\kappa}^2}{(a+1)^2 (1-b)^3 c (c+1)} & 0 \\
 0 & 0 & 0 & 0 & 0 & 0 & 0 & 0 \\
 0 & 0 & 0 & 0 & 0 & 0 & 0 & 0 \\
 \frac{\Omega_0 (1-a) (1-c) S_0S_1 \tilde{\kappa}^2 }{4 (a+1) (1-b)^3 c (c+1)} & 0 & 0 & \frac{ (b-c) (1-c)  S_0S_1\tilde{\kappa}^2}{(a+1)^2 (1-b)^3 c (c+1)} & 0 & 0 & -\frac{(1-a) (1-c)^2 (b-c) S_0^2S_1  \tilde{\kappa}^4}{2 (a+1)^3 (1-b)^4 c (c+1)} & 0 \\
 0 & 0 & \frac{(1-c) S_0S_1 \tilde{\kappa}^2}{(a+1)^2 (1-b)^2 (c+1)}  & 0 & 0 & 0 & 0 & \frac{(1-a) (1-c)^2 S_0^2S_1 \tilde{\kappa}^4}{2 (a+1)^3 (1-b)^3 (c+1)} \\
\end{array}
\right)\,,\\
     A_2&=\left(
\begin{array}{cccccccc}
 \frac{ a^2 b(b-c) (b+1) (1-c) \tilde{\kappa}^2}{c(1-b)^3 } & 0 & 0 & \frac{\Omega_0 a^2 b}{2c (1-b)^2 } & 0 & 0 & -\frac{\Omega_0 S_2 a (1-a) \tilde{\kappa}^2}{4c (a+1) (1-b)^3 } & 0 \\
 0 & 0 & 0 & 0 & 0 & 0 & 0 & 0 \\
 0 & 0 & 0 & 0 & 0 & 0 & 0 & 0 \\
 -\frac{\Omega_0a^2 b}{2 (1-b)^2 c} & 0 & 0 & -\frac{2 a^2 b^2 (1-c)}{c\left(1-a^2\right) (1-b)^2 } & 0 & 0 & \frac{S_2 a b (1-c) \tilde{\kappa}^2}{(a+1)^2 (1-b)^3 c} & 0 \\
 0 & 0 & 0 & 0 & \frac{b (1-c) \tilde{\kappa}^2}{1-b} & 0 & 0 & -\frac{\Omega_0 \left(1-a^2\right) \tilde{\kappa}^2}{4 (1-b)} \\
 0 & 0 & 0 & 0 & 0 & 0 & 0 & 0 \\
 -\frac{\Omega_0S_2 a (1-a) \tilde{\kappa}^2}{4c(a+1) (1-b)^3} & 0 & 0 & \frac{S_2 a b (1-c) \tilde{\kappa}^2}{c(a+1)^2 (b-1)^3 } & 0 & 0 & \frac{S_2^2 (1-a) (1-c) \tilde{\kappa}^4}{2  c(a+1)^3 (b-1)^4} & 0 \\
 0 & 0 & 0 & 0 & \frac{\Omega_0 \left(1-a^2\right) \tilde{\kappa}^2}{4 (1-b)} & 0 & 0 & -\frac{(b-c)\left(1-a^2\right) (b+1) (1-c) \tilde{\kappa}^4 }{2 (1-b)^2} \\
\end{array}
\right)\,,\\
    A_3&=
\left(
\begin{array}{cccccccc}
 \frac{2 b(b-c) S_3  (1-c)  \tilde{\kappa}^2}{(1-b)^3 (c+1)} & 0 & 0 & \frac{S_3 \Omega_0}{2 (1-b)^2 (c+1)} & 0 & 0 & -\frac{ \Omega_0S_3 S_4 (1-a) \tilde{\kappa}^2}{4 (a+1) (1-b)^3 (c+1)} & 0 \\
 0 & 0 & 0 & 0 & 0 & 0 & 0 & 0 \\
 0 & 0 & -\frac{S_3 (1-c)}{\left(1-a^2\right) (1-b) (c+1)} & 0 & 0 & 0 & 0 & -\frac{S_3 S_4 (1-c) \tilde{\kappa}^2}{2 (a+1)^2 (1-b)^2 (c+1)} \\
 -\frac{ \Omega_0 S_3}{2 (1-b)^2 (c+1)} & 0 & 0 & -\frac{S_3 (b+1) (1-c)}{\left(1-a^2\right) (1-b)^2 (c+1)} & 0 & 0 & \frac{S_3 S_4 (b+1) (1-c) \tilde{\kappa}^2}{2 (a+1)^2 (1-b)^3 (c+1)} & 0 \\
 0 & 0 & 0 & 0 & 0 & 0 & 0 & 0 \\
 0 & 0 & 0 & 0 & 0 & 0 & 0 & 0 \\
 -\frac{\Omega_0 S_3 S_4 (1-a)  \tilde{\kappa}^2}{4 (a+1) (1-b)^3 (c+1)} & 0 & 0 & -\frac{S_3 S_4 (b+1) (1-c) \tilde{\kappa}^2}{2 (a+1)^2 (1-b)^3 (c+1)} & 0 & 0 & \frac{S_3 S_4^2 (1-a) (b+1) (1-c) \tilde{\kappa}^4}{4 (a+1)^3 (b-1)^4 (c+1)} & 0 \\
 0 & 0 & \frac{S_3 S_4 (1-c) \tilde{\kappa}^2}{2 (a+1)^2 (1-b)^2 (c+1)} & 0 & 0 & 0 & 0 & \frac{S_3 S_4^2 (1-a) (1-c) \tilde{\kappa}^4}{4 (a+1)^3 (1-b)^3 (c+1)} \\
\end{array}
\right)\,,
\end{split}
\end{align}
\end{scriptsize}
where we introduced 
\begin{align}
    S_0&=a b+a+1\,,\\
    S_1&=-c \left(a^2+b\right)+a^2 b+c\,,\\
    S_2&=a b (a b+a+1)-c \left((a-2) (a+1) b+(a+1)^2+b^2\right)\,,\\
    S_3&=-a^2 (b+1)-b+1\,,\\
    S_4&=b (2 a+c+1)-2 (a+1) c\,.
\end{align}
All of these residue matrices $A_j$ are of rank 2. This enables us to straightforwardly employ the solution generating techniques based on the BM linear system.

\subsubsection*{Charge matrix}

We now evaluate the charge matrix associated with the monodromy matrix (\ref{ct-mono}). Since the rod structure of the black lens contains the same number of turning  points as that of the black ring, the charge matrix $Q$ takes the same functional form as in the black ring case: 
\begin{align}
    Q&=Y_{\rm flat}^{-1}\left(\sum_{j=1}^{3}A_j\right)\,.
\end{align}
When we expand it in terms of the $\mathfrak{so}(4,4)$ generators, $Q$ is expressed as
\begin{align}
\begin{split}\label{charge-lens}
    Q
    &=-\frac{1}{2}b\frac{1-c}{1-b}\tilde{\kappa}^2 \sum_{j=1}^{3}H_j
    +Q_{E_0}E_0+F_0+\frac{\Omega_0(1-a^2)\tilde{\kappa}^2}{4(1-b)}(E_{p_0}+E_{q^0})\,,\\
    Q_{E_0}&=-\frac{(1-a) (1-c) \left(b^2 (a (4 a+c+3)-c+1)+2 (a+1) b (2 a (1-c)+c+1)-4 (a+1)^2 c\right)}{4 (a+1)^2 (1-b)^2}\tilde{\kappa}^4\,.
\end{split}
\end{align}
Wne can check that the expression (\ref{charge-lens}) by using the asymptotic quantities (\ref{asym-lens}).
As in the black ring case, the charge matrix satisfies a slightly modified cubic relation
\begin{align}
    Q^3-\frac{1}{4}\Tr(Q^2)Q=q_{H}\left(H_1-\frac{1}{2}H_2-\frac{1}{2}H_3\right)\,,
\end{align}
where the trace of the square of the charge matrix $Q$ and the constant $q_{H}$ are
\begin{align}
\begin{split}
    \Tr(Q^2)&=\frac{2 (1-c)\left(\left(1-a^2\right) b (2 a (c-1)-c-1)+b^2 \left(2 a^3+a^2 (1-c)+a (2-4 c)-c+1\right)+2 (1-a) (a+1)^2 c\right)}{(a+1)^2 (1-b)^2}\tilde{\kappa}^4\,,\\
    q_{H}&=\frac{b (1-c)^2 \left(a^2 (b+1)+b-1\right) \left(a^2 (b-c)-b c+c\right)}{2 (a+1)^2 (1-b)^3}\tilde{\kappa}^6\,.
\end{split}
\end{align}

\subsection{Factorization of monodromy matrix}

We explicitly perform the factorization of the monodromy matrix (\ref{ct-mono}).
To this end, we express the residue matrices $A_j$ in terms of the eight-component vectors $a_j$ and $b_j$ satisfying the relation (\ref{rank2-mat}).
Here, we construct these matrices $A_j$ using the eigenvectors with the non-zero eigenvalues; the explicit expressions are shown as
\begin{align}
\begin{split}\label{ab-vec-lens}
    a_1^{T}&=\left(0,0,-\frac{2 (a+1) (1-b)}{(1-a) (1-c)S_0\tilde{\kappa}^2 },0,0,0,0,1\right)\,,\\
    b_1^{T}&=\left(1,0,0,-\frac{\Omega_0(1-b) }{2 b (b+1) (1-c)^2 \tilde{\kappa}^2},0,0,-\frac{ \Omega_0(1-a)S_0}{4 (a+1) b (b+1) (1-c)},0\right)\eta\,,\\ 
    a_2^{T}&=\left(0,0,0,0,1,0,0,\frac{\Omega_0\left(1-a^2\right) }{4 b (1-c)}\right)\,,\\
   b_2^{T}&=\left(-\frac{\Omega_0 a (a+1)^2 (1-b)}{2 S_2(1-c) \tilde{\kappa}^2},0,0,\frac{2 a (a+1) (1-b) b}{S_2(1-a) \tilde{\kappa}^2},0,0,1,0\right)\eta\,,\\
    a_3^{T}&=\left(0,0,-\frac{2 (a+1) (b-1)}{S_4(a-1) \tilde{\kappa}^2},0,0,0,0,1\right)\,,\\
   b_3^{T}&=\left(-\frac{\Omega_0 (a+1)^2 (1-b)}{S_4(b+1) (1-c) \tilde{\kappa}^2},0,0,\frac{2 (a+1) (1-b)}{S_4(1-a) \tilde{\kappa}^2},0,0,1,0\right)\eta\,,
\end{split}
\end{align}
and the constants $\alpha_j,\beta_j$ are given by
\begin{align}
\begin{split}
      \alpha_1&=\frac{S_1(1-a) (1-c)^2 S_0^2\tilde{\kappa}^4}{2 (a+1)^3 (1-b)^3 (c+1)}\,,\qquad \beta_1= -\frac{b (b+1) (1-c)^2 S_1\tilde{\kappa}^2}{(1-b)^3 c (c+1)}\,,\\
      \alpha_2&= -\frac{b (1-c) \tilde{\kappa}^2}{1-b}\,,\qquad 
      \beta_2=\frac{ (1-a) (1-c) S_2^2\tilde{\kappa}^4}{2 (a+1)^3 (1-b)^4 c}\,,\\
      \alpha_3&= \frac{ (1-a) (1-c)S_3S_4^2 \tilde{\kappa}^4}{4 (a+1)^3 (1-b)^3 (c+1)}\,,\qquad
      \beta_3= \frac{ (1-a) (b+1) (1-c) S_3S_4^2\tilde{\kappa}^4}{4 (a+1)^3 (1-b)^4 (c+1)}\,.
\end{split}
\end{align}
From the expressions (\ref{ab-vec-lens}) of $a_j,b_j$, we obtain the $3\times 3$ matrices $\Gamma^{(0)}$ and $\Gamma^{(a)},\Gamma^{(b)}$ as
\begin{align}
\begin{split}\label{gamma-lens}
    \Gamma_{11}^{(0)}&=-\frac{(a+1) (1-b)^2 \Omega_{0} (S_3 (1-c)-(a+1) (1-b) (c+1))}{2 S_1 b (b+1) (1-c)^3S_0\tilde{\kappa}^4}\frac{1}{\la_1\nu_1}\,,\\
 \Gamma_{12}^{(0)}&=\frac{2 (a+1)^2 (1-b)^2 c (a b+a-b+1)}{S_2 (1-a) (1-c) S_0 \tilde{\kappa}^2\la_{12}}\,,\\
  \Gamma_{13}^{(0)}&=\frac{2 (a+1)^2 (1-b)^2 (c+1)}{S_4 (1-a) (1-c)S_0\tilde{\kappa}^2 \la_{13} }\,,\\
       \Gamma_{21}^{(0)}&=-\frac{(1-b) c}{b (1-c)\la_{12}}\,,\\
     \Gamma_{22}^{(0)}&=-\frac{4a (a+1) \left(1-b^2\right)  (b-c)}{(1-a) \Omega_{0} S_2\tilde{\kappa}^2}\frac{1}{\la_2\nu_2}\,,\\
     \Gamma_{23}^{(0)}&=\frac{(a+1)^2 (1-b)^2 \Omega_{0}}{2 S_4 b (b+1) (1-c) \tilde{\kappa}^2 \la_{23}}\,,\\
     \Gamma_{31}^{(0)}&=-\frac{(a+1) (1-b)^2 (c+1) \Omega_{0}}{2 S_4 b (b+1) (1-c)^2 \tilde{\kappa}^2\la_{13}}\,,\\
     \Gamma_{32}^{(0)}&=\frac{2 (a+1)^2 (1-b)^2 (S_4-(a+1) (b-c))}{S_2 S_4 (1-a) \tilde{\kappa}^2 \la_{23}}\,,\\
     \Gamma_{33}^{(0)}&=\frac{4(a+1)^2 (1-b)^2 \left(2 S_1-(a+1) (1-b) (c+1)\right)}{S_3 S_4^2 (1-a) (1-c)\tilde{\kappa}^4}\frac{1}{\la_3\nu_3}\,,
\end{split}
\end{align}
and
\begin{align}
    \Gamma^{(a)}&=\Gamma^{(b)}=0_{3\times 3}\,.
\end{align}
The matrix $X_+(\lambda)$ can be obtained from the relations (\ref{xp-ex}) and (\ref{c-def}), and hence $X_-(\lambda)=X_+(-1/\lambda)$ is also obtained. Hence, we can see that the monodromy matrix $\cM_{\rm CT}(w)$ can be factorized
\begin{align}
    \cM_{\rm CT}(w(\la,z,\rho))=X_-(\la,z,\rho)M_{\rm CT}(z,\rho)X_+(\la,z,\rho)\,.
\end{align}
Therefore, the monodromy matrix (\ref{ct-mono}) describes the 5D rotating black lens solution (\ref{rrrlens-f}) constructed by Chen and Teo. In particular, the monodromy matrix (\ref{ct-mono}) can be regarded as a unified matrix that captures all three types of black hole solutions with different horizon topologies discussed in this paper.

\subsubsection*{Conformal factor}

Finally, we close this section by computing the conformal factor $e^{2\nu}$.
Since $\Gamma^{(a)}=\Gamma^{(b)}=0_{3\times 3}$, we can again use the simplified formula (\ref{conf}).
Using the expression (\ref{gamma-lens}) of $\Gamma^{(0)}$, the formula (\ref{conf}) takes the form
\begin{align}\label{conf-lens-com}
   k_{\rm BM}\prod_{j=1}^{3}(\la_j\nu_j)\,{\rm det}(\Gamma^{(0)})
    &=k_{\rm BM}\left(-\frac{4 (a+1)^5 (1-b)^6 c (c+1)^2}{ (1-a)^3 b (b+1) (1-c)^4  S_0S_1S_2S_3 S_4^2\tilde{\kappa}^{12}}\right)\no\\
    &\quad\times \frac{(u-v)\left(F(v,u)-F(u,v)-2J(u,v)\right)}{(u+v+c(1+uv))(2+c(1+u+v-uv))(2+c(-1+u+v+uv))}\,.
\end{align}
This precisely reproduces the conformal factor (\ref{conf-lens}) for the 5D rotating black lens solution by taking the constant $k_{\rm BM}$ as
\begin{align}
    &k_{\rm BM}=-\frac{(1-a)^2 b(1+b) (1-c)^4S_0S_1 S_2 S_3 S_4^2  \tilde{\kappa}^{10}}{8c (1+a)^6(1-b)^9(1+c)^2}\,.
\end{align}

\newpage

\section{Conclusion and discussion}\label{sec:dis}

In this work, we had considered a solution-generating technique based on the BM linear system for constructing asymptotically flat black hole solutions of five-dimensional vacuum Einstein theory. 
We had focused on three classes of asymptotically flat, singly rotating black hole solutions: (i) the Myers-Perry black hole~\cite{Myers:1986un}, (ii) the Emparan-
Reall black ring~\cite{Emparan:2004wy}, and (iii) the Chen-Teo black lens~\cite{Chen:2008fa}, each characterized by a distinct event-horizon topology.
By employing the angular coordinate system introduced in~\cite{Giusto:2007fx}, we had presented the corresponding monodromy matrices for the $SO(4,4)$ Geroch group, ensuring that all entries remained finite. 
Furthermore, we had confirmed that these three black hole solutions could be precisely reconstructed by factorizing their respective monodromy matrices. 
This had shown that the solution-generating technique remained effective well beyond the spherical horizon case.
In addition, we had obtained the universal expression (\ref{Q-gen}) for the monodromy matrices associated with the asymptotic behavior of five-dimensional asymptotically flat vacuum solutions described in~\cite{Harmark:2004rm}. 
We had explicitly demonstrated that the constant matrix $Y_{\rm flat}$ and a part of the charge matrix $Q$ defined in~(\ref{mono-asym}) were uniquely determined by the conditions of asymptotic flatness together with the asymptotic conserved quantities, namely the mass and angular momentum.

A natural extension of this work would be to examine the doubly rotating case with two nonzero angular momenta. 
A prominent example is the Pomeransky-Sen'kov black ring~\cite{Pomeransky:2006bd} and its unbalanced generalization~\cite{Morisawa:2007di,Chen:2011jb} and the rotating black lens with two non-zero angular momenta~\cite{Tomizawa:2019acu}. 
Beyond this, extending the monodromy-matrix framework to incorporate non-BPS black hole solutions with nontrivial $U(1)$ gauge fields and scalar fields in five-dimensional minimal supergravity and $U(1)^3$ supergravity is essential for deepening our understanding of how matter fields are encoded in the monodromy data. 
In such theories, there exist black hole solutions whose horizon cross sections are spherical, but whose domains of outer communication are topologically nontrivial~\cite{Kunduri:2014iga,Suzuki:2023nqf,Suzuki:2024abu}.
Another important direction is to clarify how the rod structure of a solution is reflected in the residue matrices of the monodromy matrix. 
This would shed light on the way in which the topology (of the horizon cross section and domain of outer communication) of the underlying geometry is captured algebraically. 
As emphasized in the introduction, the uniqueness theorems~\cite{Morisawa:2004tc,Hollands:2007aj,Tomizawa:2009ua,Tomizawa:2009tb} for five-dimensional asymptotically flat, stationary and bi-axisymmetric black holes suggest that the monodromy matrix should be uniquely fixed by the asymptotic charges together with the rod structure.
Pursuing these directions would not only help to clarify the structure of the moduli space of solutions to the Einstein equations, but would also open a path toward more powerful solution-generating techniques. Such methods could enable the systematic construction of new black hole spacetimes with richer topological and matter-field content, including regular black lens solutions, solutions with topologically trivial domains of outer communication, and the solutions describing multi-black hole or black-object configurations.

As another avenue for future investigation, it would be valuable to study the implications of the nilpotent condition of the charge matrix $Q$ in the doubly rotating case. We have shown that, for the singly rotating Myers-Perry black hole, the nilpotency of the $SO(4,4)$ charge matrix $Q$ itself encodes the extremality conditions, in contrast to the $SL(3,\mathbb{R})$ case. It is natural to expect that a similar feature should hold for the doubly rotating Myers-Perry black hole.
Even more intriguingly, for black rings it would be important to examine whether the nilpotent condition of $Q$ can likewise characterize extremality, as it does for spherical horizons. If this turns out to be the case, the charge matrix $Q$ could serve as a powerful diagnostic tool for classifying and constructing new extremal solutions beyond the spherical horizon class.

\section*{Acknowledgements}

S.T.\ was supported by JSPS KAKENHI Grant Number 21K03560.

\newpage

\appendix

\section{Asymptotic behavior of scalar fields }

In this appendix, we provide a detailed computation of the asymptotic behavior of the scalar fields parameterizing the symmetric coset $SO(2,2)\times SO(2,2)\backslash SO(4,4)$, corresponding to the asymptotic form of the five-dimensional metric (\ref{metric-asy}) at spatial infinity.

It is well established that the moduli space of five-dimensional asymptotically flat axisymmetric solutions can be realized as a symmetric coset space after performing dimensional reduction to three dimensions. As in \cite{Sakamoto:2025jtn}, we perform a reduction in two step i.e. first along the time direction, and then along the $\psi$ direction, using the Kaluza-Klein ansatz given below:
\begin{align}
\begin{split}
       ds_5^2&=-f^2(dt+\check{A}^0)^2+f^{-1}ds_4^2\,,\\
    A^{I}&=\chi^I(dt+\check{A}^0)+\check{A}^I\,,
\end{split}
\end{align}
and
\begin{align}
\begin{split}
    ds_4^2&=e^{2U}(d\psi+\omega_3)^2+e^{-2U}ds_3^2\,,\\
     \check{A}^{\Lambda}&=\zeta^{\Lambda}(d\psi+\omega_3)+\hat{A}^{\Lambda}\,,
\end{split}
\end{align}
where the 3D base space is described by the metric
\begin{align}
      ds_3^2&=e^{2\nu}(d\rho^2+dz^2)+\rho^2 d\phi^2\,.
\end{align}
Since we consider vacuum solutions, the scalar fields $\{e^{2U}, x^I, y^I,\zeta^{\Lambda}\}$ are
\begin{align}
\begin{split}\label{scalar-rep}
    e^{2U}&=\frac{\left(g_{t\tilde{\phi}}-g_{t\tilde{\psi}}\right)^2-g_{tt}\left(g_{\tilde{\phi}\tilde{\phi}}-2g_{\tilde{\phi}\tilde{\psi}}+g_{\tilde{\psi}\tilde{\psi}}\right)}{4\sqrt{-g_{tt}}}\,,\\
    x^I&=-\chi^I=0\,,\qquad y^I=fh^I=\sqrt{-g_{tt}}\,,\\
    \zeta^0&=\frac{g_{t\tilde{\phi}}-g_{t\tilde{\psi}}}{2g_{tt}}\,,\qquad \zeta^I=0\,,
\end{split}
\end{align}
where we set $h^I=1$\footnote{For example, a five-dimensional asymptotically flat vacuum solution can be embedded into eleven-dimensional supergravity as a direct product of the form (5D vacuum solution)$\times (T^2)^3$. 
In this embedding, the scalar fields $h^I$ are interpreted as moduli fields that describe the volumes of the individual two-tori.}.
The one-forms $\hat{A}^{\Lambda}$ and $\omega_3$ are
\begin{align}
 \hat{A}^{0}&=-\frac{g_{t\tilde{\phi}}\left(g_{\tilde{\psi}\tilde{\psi}}-g_{\tilde{\phi}\tilde{\psi}}\right)+g_{t\tilde{\psi}}\left(g_{\tilde{\phi}\tilde{\phi}}-g_{\tilde{\phi}\tilde{\psi}} \right)  }{4\sqrt{-g_{tt}}e^{2U} }d\phi\,,\qquad  \hat{A}^{I}=0\,,\\
 \omega_3&=\frac{g_{t\tilde{\phi}}^2-g_{t\tilde{\psi}}^2-g_{tt}\left(g_{\tilde{\phi}\tilde{\phi}}-g_{\tilde{\psi}\tilde{\psi}} \right)  }{4\sqrt{-g_{tt}}e^{2U} }d\phi\,.
\end{align}

Now we compute the asymptotic behavior of the scalar fields under the limit $\sqrt{\rho^2+z^2}\to \infty$ with $z/\sqrt{\rho^2+z^2}$ fixed.
By substituting the asymptotic behavior (\ref{metric-asy}) of the metric into (\ref{scalar-rep}), we have 
\begin{align}
\begin{split}\label{scalar-asy}
    e^{2U}&\simeq\sqrt{\rho^2+z^2}
    \biggl(1-\frac{\eta}{3\pi}\frac{z}{\rho^2+z^2}+\cO\left(\frac{1}{\rho^2+z^2}\right)\biggr)\,,\\
      y^I&\simeq1-\frac{M}{3\pi }\frac{1}{\sqrt{\rho^2+z^2}}+\cO\left(\frac{1}{\rho^2+z^2}\right)\,,\\
    \zeta^0&\simeq \frac{J_{1}}{4\pi }\frac{\sqrt{\rho^2+z^2}-z}{\rho^2+z^2}-\frac{J_{2}}{4\pi }\frac{\sqrt{\rho^2+z^2}+z}{\rho^2+z^2}+\cO\left(\frac{1}{\rho^2+z^2}\right)\,,
\end{split}
\end{align}
and
\begin{align}
    \hat{A}^0&\simeq \left(\frac{J_{1}+J_{2}}{4\pi }\frac{\rho^2}{(\rho^2+z^2)^{\frac{3}{2}}}+\cO\left(\frac{1}{\rho^2+z^2}\right)\right)d\phi\,,\\
    \omega_3&\simeq \left(-\frac{z}{\sqrt{\rho^2+z^2}} +\frac{\eta}{3\pi}\frac{\rho^2}{\sqrt{\rho^2+z^2}}+\cO\left(\frac{1}{\rho^2+z^2}\right)\right)d\phi\,.
\end{align}
We introduce the field strengths $\hat{F}_2^{\Lambda}$ and $\hat{F}_2$ for the one-form fields $\hat{A}^{\Lambda}$ and $\omega_3$
\begin{align}
   \hat{F}_2^{\Lambda}=d\hat{A}^{\Lambda} \,,\qquad \hat{F}_2=d\omega_3\,.
\end{align}
In three-dimensional space, the field strengths $\hat{F}_2^{\Lambda}$ and $\hat{F}_2$ are dualized into scalar fields $\tilde{\zeta}_{\Lambda}$ and $\sigma$ via Hodge duality:
\begin{align}
\begin{split}\label{dual-eq}
    -d\tilde{\zeta}_{\Lambda}&=e^{2U}({\rm Im}\,N)_{\Lambda \Sigma}\star_{3}(\hat{F}_2^{\Sigma}+\zeta^{\Sigma}\hat{F}_2)+({\rm Re}\,N)_{\Lambda\Sigma}d\zeta^{\Sigma}\,,\\
   -d\sigma&= -2e^{4U}\star_{3}\hat{F}_2+\tilde{\zeta}_{\Lambda}d\zeta^{\Lambda}-\zeta^{\Lambda}d\tilde{\zeta}_{\Lambda} \,,
\end{split}
\end{align}
where $\star_3$ is the Hodge star operator relative to the three-dimensional base space $ds_3^2$.
In our set up, the symmetric matrix $N_{\Lambda\Sigma}$ takes the expression
\begin{align}
    {\rm Re}\,N=0_{4\times 4}\,,\qquad {\rm Im}\,N={\rm diag}(y^3,y,y,y)\,.
\end{align}
Hence, the duality relation of the twist potential $\tilde{\zeta}_{0}$ reduces to
\begin{align}
     d\tilde{\zeta}_{0}&=-e^{2U}({\rm Im}\,N)_{00}\star_{3}(\hat{F}_2^{0}+\zeta^{0}\hat{F}_2)\no\\
     &\simeq-\frac{1}{\pi}\left(\frac{J_{1}-J_{2}}{4}\frac{1}{(\rho^2+z^2)^{\frac{1}{2}}}+\frac{J_1+J_2}{2}\frac{z}{\rho^2+z^2}\right)\frac{\rho}{\rho^2+z^2}d\rho\no\\
     &\quad-\frac{1}{\pi}\left(\frac{J_{1}-J_{2}}{4}\frac{z}{(\rho^2+z^2)^{\frac{1}{2}}}-\frac{J_1+J_2}{2}\frac{\rho^2-z^2}{\rho^2+z^2}\right)\frac{1}{\rho^2+z^2}dz\,.
\end{align}
By solving the equation, we obtain
\begin{align}\label{tzeta-asy}
    \tilde{\zeta}_0&\simeq \frac{J_{1}}{4\pi }\frac{\sqrt{\rho^2+z^2}+z}{\rho^2+z^2}-\frac{J_{2}}{4\pi }\frac{\sqrt{\rho^2+z^2}-z}{\rho^2+z^2}+\cO\left(\frac{1}{\rho^2+z^2}\right)\,.
\end{align}
The duality relation (\ref{dual-eq}) of the twist potential $\sigma$ becomes
\begin{align}
    d\sigma\simeq \left(\frac{2\rho}{\sqrt{\rho^2+z^2}}+\frac{2\eta}{3\pi}\frac{\rho z}{(\rho^2+z^2)^{\frac{3}{2}}}\right)d\rho+\left(\frac{2z}{\sqrt{\rho^2+z^2}}-\frac{2\eta}{3\pi}\frac{\rho^2}{(\rho^2+z^2)^{\frac{3}{2}}}\right)dz\,.
\end{align}
By integrating over $\rho$ and $z$, we obtain
\begin{align}\label{sigma-asy}
   \sigma&=2\sqrt{\rho^2+z^2}\left(1-\frac{\eta}{3\pi}\frac{z}{\rho^2+z^2}+\cO\left(\frac{1}{\rho^2+z^2}\right)\right)\,.
\end{align}

For completeness, we finally compute the asymptotic behaviour of the monodromy matrix in the large spectral parameter region. From the parameterization of (\ref{iwasawa-rep}) of an $SO(4,4)$ element with the vacuum configuration (\ref{vacuum-scalar}), the coset matrix $M(z,\rho)$ is taken as the expression
\begin{align}
  M=  \left(
\begin{array}{cccccccc}
 \frac{1}{y^2}-\frac{y (\zeta^{0})^2}{e^{2U}} & 0 & 0 & \frac{y \zeta^{0}}{e^{2U}} & 0 & 0 & \frac{y \zeta^{0} (\zeta^{0} \tilde{\zeta}_0-\sigma)}{2 e^{2U}}-\frac{\tilde{\zeta}_0}{y^2} & 0 \\
 0 & 1 & 0 & 0 & 0 & 0 & 0 & 0 \\
 0 & 0 & \frac{1}{y e^{2U}} & 0 & \frac{\tilde{\zeta}_0}{y e^{2U}} & 0 & 0 & \frac{\sigma+\zeta^{0} \tilde{\zeta}_0}{2 y e^{2U}} \\
 -\frac{y \zeta^{0}}{e^{2U}} & 0 & 0 & \frac{y}{e^{2U}} & 0 & 0 & -\frac{y (\sigma-\zeta^{0} \tilde{\zeta}_0)}{2 e^{2U}} & 0 \\
 0 & 0 & \frac{\tilde{\zeta}_0}{y e^{2U}} & 0 & y^2+\frac{\tilde{\zeta}_0^2}{y e^{2U}} & 0 & 0 & y^2 \zeta^{0}+\frac{\tilde{\zeta}_0 (\sigma+\zeta^{0} \tilde{\zeta}_0)}{2 y e^{2U}} \\
 0 & 0 & 0 & 0 & 0 & 1 & 0 & 0 \\
 \frac{y \zeta^{0} (\zeta^{0} \tilde{\zeta}_0-\sigma)}{2 e^{2U}}-\frac{\tilde{\zeta}_0}{y^2} & 0 & 0 & \frac{y (\sigma-\zeta^{0} \tilde{\zeta}_0)}{2 e^{2U}} & 0 & 0 & m_{77} & 0 \\
 0 & 0 & -\frac{\sigma+\zeta^{0} \tilde{\zeta}_0}{2 y e^{2U}} & 0 & -y^2 \zeta^{0}-\frac{\tilde{\zeta}_0 (\sigma+\zeta^{0} \tilde{\zeta}_0)}{2 y e^{2U}} & 0 & 0 & m_{88}\\
\end{array}
\right)\,,
\end{align}
where we denoted all $y^I$ by $y$ and defined as
\begin{align}
    m_{77}&=y e^{2U}+\frac{\tilde{\zeta}_0^2}{y^2}-\frac{y (\sigma-\zeta^0 \tilde{\zeta}_0)^2}{4 e^{2U}}\,,\\
    m_{88}&=\frac{e^{2U}}{y}-y^2 (\zeta^0)^2-\frac{(\sigma+\zeta^0 \tilde{\zeta}_0)^2}{4 y e^{2U}}\,.
\end{align}
Using the asymptotic behaviors (\ref{scalar-asy}), (\ref{tzeta-asy}) and (\ref{sigma-asy}) of the scalar fields computed above, we evaluate the monodromy matrix from the relation (\ref{sub-rule}) and then obtain 
\begin{align}
\cM(w)\sim Y_{\text{flat}}\left(1+\frac{Q}{w}\right)=\left(
\begin{array}{cccccccc}
 1-\frac{2 M}{3 \pi  w} & 0 & 0 & 0 & 0 & 0 & \frac{J_1-J_2}{2 \pi  w} & 0 \\
 0 & 1 & 0 & 0 & 0 & 0 & 0 & 0 \\
 0 & 0 & -\frac{1}{w} & 0 & 0 & 0 & 0 &1 -\frac{M}{3 \pi  w} \\
 0 & 0 & 0 & -\frac{1}{w} & 0 & 0 & -1-\frac{M}{3 \pi  w} & 0 \\
 0 & 0 & 0 & 0 & 1+\frac{2 M}{3 \pi  w} & 0 & 0 & -\frac{J_1-J_2}{2 \pi  w} \\
 0 & 0 & 0 & 0 & 0 & 1 & 0 & 0 \\
 \frac{J_1-J_2}{2 \pi  w} & 0 & 0 & 1+\frac{M}{3 \pi  w} & 0 & 0 & 0 & 0 \\
 0 & 0 &-1+ \frac{M}{3 \pi  w} & 0 & \frac{J_1-J_2}{2 \pi  w} & 0 & 0 & 0 \\
\end{array}
\right)\,.
\end{align}
This result leads to the $\mathfrak{so}(4,4)$ basis expansion (\ref{Q-gen}) of the charge matrix $Q$ with $Q_{E_0}=0$\,. 
The coefficient of $E_0$ can be computed from the components $m_{77}$ and $m_{88}$.
As can be seen from their explicit expressions, obtaining a non-vanishing value is necessary for incorporating the next-order contributions from $e^{2U}$ and $\sigma$.

\section{Explicit expression of twist potential}\label{sec:twist}

In this appendix, we give an explicit expression of the twist potential $\sigma$ in the $C$-metric
\begin{align}
    \sigma(u,v)&=\frac{(1-a)\tilde{\kappa}^2}{4(1+a)(u-v)H(u,v)H(v,u)}\sigma_0(u,v)\,.
\end{align}
The numerator $\sigma_0(u,v)$ is a symmetric polynomial of $u$ and $v$ taking the form
\begin{align}
    \sigma_0(u,v)=\sum_{m,n=0}^{5}\sigma_0^{m,n}u^mv^n\,.
\end{align}
The coefficients $\sigma_0^{m,n}$ are
\begin{align}
    \sigma_0^{5,5}&= -a^4c^3 (b-c)^2(1-a) (1+a)^3 (1-b^2)^2\,,\\
    \sigma_0^{4,5}&=- a^4 c^2 (b-c) (1+a)^2 (1-b^2)\Bigl(b^3 \left(a^2 (c+2)-2 c^2+c-2\right)\no\\
    &\quad+b^2 \left(-a^2 \left(c^2+3 c-1\right)+3 c^2-c+1\right)+b \left(a^2 \left(c^2-3 c-1\right)+c^2+c+1\right)+\left(a^2-1\right) c (2 c+1)\Bigr)\,,\\
      \sigma_0^{3,5}&=-2 a^3c (b-c) (1+a) (1+b)  \Bigl(a^4 \left(b^4 (c-2) c+b^3 \left(-c^3+2 c^2+1\right)-b^2 (c-1)+b (c-3) c-(c-2) c^2\right)\no\\
      &\quad+a^3c \left(b^2-1\right)  \left(b^2 (2 c-3)+b \left(-2 c^2+3 c+1\right)-c\right)\no\\
      &\quad-a^2 (b-1) \left(b^3 \left(c^3-3 c^2+2 c-1\right)+b^2 c^2 (3-2 c)+b c \left(2 c^2-4 c+3\right)+c^2 (2 c-3)\right)
      \no\\
      &\quad-a (b-1)^2 \left(b^2 \left(2 c^3-5 c^2+3 c-1\right)+b (c-1) c^2+c^2\right)+(b-1)^2 (c-1) c \left(b^2 c-b c+b-c\right)\Bigr)\,,\\
     \sigma_0^{2,5}&=-2 a^3 (b+1) (b-c) \Bigl(a^5 (b+1) c \left(b^3 c^2-b^2 \left(c^3+2 c^2+3 c-3\right)+b c \left(2 c^2+7 c-6\right)+c^2 (3-4 c)\right)\no\\
     &\quad+a^4 \bigl(b^4 \left(2 c^3+4 c^2-5 c+1\right)+b^3 \left(-2 c^4-9 c^3+9 c^2-3 c+1\right)+b^2 c \left(5 c^3-6 c^2-c+2\right)\no\\
     &\quad+2 b c^2 \left(c^2+4 c-3\right)+c^3 (3-5 c)\bigr)\no\\
     &\quad-a^3 (b-1) (c-1) c \left(b^3 \left(c^2-1\right)+b^2 (2-8 c)+6 b (c-1) c+6 c^2\right)\no\\
     &\quad-a^2 \bigl(b^4 \left(2 c^4+8 c^3-13 c^2+6 c-1\right)-b^3 \left(11 c^4+c^3-14 c^2+7 c-1\right)\no\\
     &\quad+b^2 c \left(8 c^3-13 c^2+4 c+1\right)+b c^2 \left(9 c^2-5\right)+2 c^3 (3-4 c)\bigr)\no\\
     &\quad-a c (b-1)^2 \left(b^2 \left(c^3-c+1\right)+2 b c \left(2 c^2-3 c+1\right)+c^2 (2 c-3)\right)\no\\
     &\quad+c(b-1)^2 (c-1)  \left(b^2 \left(4 c^2-2 c+1\right)-b (c-1) c-3 c^2\right)\Bigr)\,,\\
     \sigma_0^{1,5}&=a^3c (b-c) (b+1)  \Bigl(a^5 (b+1) \left(b^3 (c-4) c-b^2 \left(c^3-4 c^2-12 c+6\right)+b c (10-19 c)+c^2 (7 c-4)\right)\no\\
     &\quad+2 a^4 (b+1) \left(b^3 \left(2 c^2-9 c+4\right)+b^2 \left(-2 c^3+12 c^2+2 c-3\right)+b c \left(-3 c^2-10 c+4\right)+c^2 (4 c-1)\right)\no\\
     &\quad-2 a^3 (b-1) (c-1) \left(b^3 \left(c^2-4 c+1\right)+6 b^2 c+2 b c (3-2 c)-6 c^2\right)\no\\
     &\quad-2 a^2 (b-1) \left(b^3 \left(3 c^3-17 c^2+13 c-2\right)+b^2 \left(6 c^3-c^2-3 c+1\right)+b c \left(c^2+7 c-5\right)+c^2 (4-7 c)\right)\no\\
     &\quad-a (b-1)^2 \left(b^2 \left(2 c^3-9 c^2+6 c-2\right)+b c \left(-3 c^2+c+2\right)+c^2 (8-5 c)\right)\no\\
     &\quad-2 (b-1)^2 (c-1) c \left(3 b^2 c+b (c-1)-3 c\right)\Bigr)\,,\\
      \sigma_0^{0,5}&=a^3 (b+1) (b-c) \Bigl(a^5 (b+1) c \left(b^3 (c-2) c+b^2 \left(-c^3+c^2+5 c-2\right)+b c \left(c^2-7 c+3\right)+c^2 (2 c-1)\right)\no\\
      &\quad+2 a^4 \left(b^4 \left(c^3-2 c^2-c+1\right)+b^3 \left(-c^4+3 c^3+2 c^2-3 c+1\right)-b^2 c \left(c^3+2 c^2-5 c+2\right)+b c^2 \left(c^2-4 c+1\right)+c^4\right)\no\\
      &\quad+2 a^3 \left(b^2-3 b+2\right) (c-1) c \left(b^2-b (c-1) c-c^2\right)\no\\
      &\quad-2 a^2 \Bigl(b^4 \left(c^3-7 c^2+6 c-1\right)+b^3 \left(2 c^4-3 c^3+9 c^2-7 c+1\right)\no\\
 &\quad-b^2 (c-1)^2 c (2 c-1)-2 b c^2 \left(c^2+c-1\right)+c^3 (2 c-1)\Bigr)\no\\
      &\quad-a (b-1)^2 c \left(b^2 \left(c^2-4 c+2\right)+b c \left(c^2-4 c+3\right)+c^2 (3-2 c)\right)\no\\
 &\quad-2 (b-1)^2 (c-1) c \left(b^2 (2 c-1)+b (c-1) c-c^2\right)\Bigr)\,,
\end{align}
\begin{align}
    \sigma_0^{4,4}&=a^3 c (a+1) (b+1) (b-c) \Bigl(a^4 (b+1) \Bigl(b^3 \left(c^2-2 c+6\right)-b^2 \left(c^3-4 c^2+16 c+2\right)\no\\
 &\quad+b c \left(-2 c^2+11 c+6\right)-c^2 (c+4)\Bigr)\no\\
    &\quad-a^3 \left(b^2-1\right) \left(b^2 \left(3 c^2-6 c-2\right)+b c \left(-3 c^2+7 c+6\right)-c^2 (c+4)\right)\no\\
    &\quad-a^2 (b-1) \left(b^3 \left(4 c^3-3 c^2+6 c-2\right)+b^2 \left(11 c^3-30 c^2+16 c-2\right)+b c \left(-4 c^2+c-2\right)+c^2 (c+4)\right)\no\\
    &\quad-a (b-1)^2 \left(b^2 \left(9 c^2-6 c+2\right)+b c \left(3 c^2-5 c+2\right)-c^2 (c+4)\right)-4 (b-1)^2 b^2 (c-1)^2 c\Bigr)\,,\\
    \sigma_0^{3,4}&=-2 a^2 \Bigl(a^6 (b+1)^2 (b-c)^2 \left(b^2 \left(c^3+7 c^2-2 c-1\right)+b c \left(-9 c^2-5 c+4\right)+c^2 (8 c-3)\right)\no\\
    &\quad+a^5 (b+1)^2 (b-c)^2 \left(b^2 \left(10 c^3+4 c^2-3 c-1\right)+b c \left(-21 c^2-4 c+5\right)+c^2 (11 c-1)\right)\no\\
    &\quad-a^4 c\left(1-b^2\right) (1-c)  \left(b^4 (3 c-7)+b^3 \left(-13 c^2+5 c+2\right)+b^2 c \left(10 c^2+5 c-1\right)-3 b (c-3) c^2-10 c^3\right)\no\\
    &\quad-a^3 \left(b^2-1\right) \Bigl(b^4 \left(16 c^4-18 c^3+11 c^2+1\right)+b^3 c \left(-16 c^4+c^3-8 c^2+9 c-6\right)\no\\
    &\quad+b^2 c^2 \left(17 c^3-18 c^2+9 c-8\right)+b c^3 \left(15 c^2-2 c+7\right)+2 c^4 (3-8 c)\Bigr)\no\\
    &\quad-a^2 (1-b)^2 \Bigl(b^4 \left(3 c^4-2 c^3+5 c-1\right)+b^3 c \left(5 c^4-9 c^3+4\right)+b^2 c^2 \left(-5 c^3+17 c^2-28 c+6\right)\no\\
    &\quad+b c^3 \left(-9 c^2+25 c-16\right)+c^4 (7-2 c)\Bigr)\no\\
    &\quad-a c (1-b)^2 (1-c) \left(b^4 \left(8 c^3-14 c^2+14 c-3\right)-b^3 (1-c)^2 (2 c-1)+b^2 c \left(-11 c^2+7 c-6\right)+5 c^3\right)\no\\
    &\quad+3 bc^2(b-c)(1-b)^3  (b+1) (1-c)^2  \Bigr)\,,\\
    \sigma_0^{4,2}&=-2 a^2 \Bigl(a^6 (b+1)^2 (b-c)^2 \left(b^2 \left(c^3+5 c^2+12 c-3\right)-6 b c \left(c^2+5 c-1\right)+c^2 (17 c-2)\right)\no\\
    &\quad+a^5 (b+1)^2 c (b-c)^2 \left(b^2 \left(3 c^2+29 c-2\right)-b \left(25 c^2+34 c+1\right)+21 c^2+7 c+2\right)\no\\
    &\quad-a^4 \left(b^2-1\right) (c-1) \Bigl(b^4 \left(2 c^3+3 c^2+15 c-6\right)-b^3 c \left(2 c^3+5 c^2+55 c-18\right)\no\\
    &\quad+b^2 c \left(2 c^3+53 c^2-41 c+2\right)+b c^2 \left(-13 c^2+55 c+2\right)-2 c^3 (13 c+2)\Bigr)\no\\
    &\quad-a^3c \left(b^2-1\right)  \Bigl(b^4 \left(5 c^3+52 c^2-41 c+14\right)+b^3 \left(-5 c^4-75 c^3+20 c^2+3 c-3\right)\no\\
    &\quad+b^2 \left(23 c^4+2 c^3-37 c^2+22 c-10\right)+b c \left(19 c^3+54 c^2-23 c+10\right)+2 c^3 (2-17 c)\Bigr)\no\\
    &\quad-a^2 (b-1)^2 \Bigl(b^4 \left(5 c^4-2 c^3+16 c^2-7 c+3\right)+b^3 c^2 \left(-5 c^3+33 c^2-59 c+31\right)\no\\
    &\quad+b^2 c \left(-7 c^4+15 c^3-31 c^2-11 c+4\right)-4 b c^2 \left(5 c^3-10 c^2+3 c+2\right)+c^3 \left(-9 c^2+20 c+4\right)\Bigr)\no\\
    &\quad+a c (b-1)^2 (c-1) \Bigl(b^4 \left(23 c^2-18 c+10\right)+b^3 \left(c^3-10 c^2+15 c-6\right)\no\\
    &\quad-b^2 \left(20 c^3+5 c^2+c+4\right)+2 b c \left(2 c^2-3 c+1\right)+c^2 (13 c+2)\Bigr)\no\\
    &\quad-(b-1)^3 b (b+1) (c-1)^2 c (3 c+2) (b-c)\Bigr)\,,\\
    \sigma_0^{4,1}&=a \Bigl(a^7 (b+1)^2 (b-c)^2 \left(b^2 \left(c^3+4 c^2-36 c+6\right)+b \left(-7 c^2+65 c-8\right) c-26 c^3+c^2\right)\no\\
    &\quad+2 a^6 (b+1)^2 (b-c)^2 \left(b^2 \left(7 c^3-24 c^2-13 c+5\right)+b \left(8 c^3+37 c^2+3 c+2\right)-c \left(15 c^2+6 c+4\right)\right)\no\\
    &\quad-2 a^5 \left(b^2-1\right) (c-1) \Bigl(-\left(b^4 \left(c^3-3 c^2+27 c-4\right)\right)+b^3 \left(c^4-8 c^3+76 c^2-22 c+4\right)\no\\
    &\quad+b^2 c \left(5 c^3-73 c^2+63 c-4\right)+b c^2 \left(24 c^2-67 c-8\right)+2 c^3 (11 c+4)\Bigr)\no\\
    &\quad-2 a^4 \left(b^2-1\right) \Bigl(b^4 \left(4 c^4-35 c^3-21 c^2+34 c-7\right)+b^3 \left(-4 c^5+61 c^4+19 c^3-36 c^2+14 c-4\right)\no\\
    &\quad+b^2 c \left(-26 c^4-6 c^3+75 c^2-57 c+14\right)+b c^2 \left(8 c^3-99 c^2+51 c-10\right)+c^4 (26 c-1)\Bigr)\no\\
    &\quad+a^3 (b-1)^2 \Bigl(b^4 \left(4 c^4-13 c^3+58 c^2-26 c+2\right)+b^3 \left(-4 c^5+52 c^4-131 c^3+131 c^2-56 c+8\right)\no\\
    &\quad+b^2 c \left(-39 c^4+156 c^3-252 c^2+109 c-24\right)+b c^3 \left(-35 c^2+69 c-34\right)+c^3 \left(-18 c^2+27 c+16\right)\Bigr)\no\\
    &\quad-2 a^2 (b-1)^2 (c-1) \Bigl(b^4 \left(14 c^3+8 c^2-9 c+2\right)+b^3 \left(-14 c^4+15 c^3+10 c^2-13 c+2\right)\no\\
    &\quad+b^2 c \left(c^3-63 c^2+48 c-16\right)+2 b c^2 \left(7 c^2-12 c+5\right)+c^3 (11 c+4)\Bigr)\no\\
    &\quad+2 a b (b-1)^3 (b+1) (c-1)^2 c^2 (b-c)+4 (b-1)^3 b (b+1) (c-1)^3 c (b-c)\Bigr)\,,\\
    \sigma_0^{0,4}&=a \Bigl(a^7 (b+1)^2 (b-c)^2 \left(b^2 \left(c^3+4 c^2-14 c+2\right)-2 b \left(3 c^2-11 c+1\right) c-7 c^3\right)\no\\
    &\quad+2 a^6 (b+1)^2 (b-c)^2 \left(b^2 \left(2 c^3+c^2-16 c+6\right)-b \left(c^3-14 c^2+c-2\right)-c \left(4 c^2+c+2\right)\right)\no\\
    &\quad-2 a^5 \left(b^2-1\right) (c-1) \Bigl(b^4 \left(c^2-7 c-2\right)+b^3 \left(-3 c^3+25 c^2-8 c+4\right)\no\\
    &\quad+b^2 c \left(2 c^3-29 c^2+27 c-2\right)+b c^2 \left(11 c^2-23 c-6\right)+2 c^3 (3 c+2)\Bigr)\no\\
    &\quad-2 a^4 \left(b^2-1\right) \Bigl(b^4 \left(c^3-c^2-23 c+16\right) c-b^3 \left(c^5-3 c^4-52 c^3+71 c^2-43 c+12\right)\no\\
    &\quad-2 b^2 \left(c^4+19 c^3-44 c^2+30 c-6\right) c+b \left(9 c^2-40 c+17\right) c^3+7 c^5\Bigr)\no\\
    &\quad+a^3 (b-1)^2 \Bigl(b^4 \left(2 c^4-11 c^3+22 c^2-6\right)-2 b^3 \left(c^5-10 c^4+22 c^3-16 c^2-c+4\right)\no\\
    &\quad+b^2 c^2 \left(-9 c^3+40 c^2-59 c+14\right)-2 b c^2 \left(9 c^3-24 c^2+23 c-8\right)+c^3 \left(-5 c^2+4 c+8\right)\Bigr)\no\\
    &\quad-2 a^2 (b-1)^2 (c-1) \Bigl(-\left(b^4 \left(c^3-18 c^2+14 c-2\right)\right)+b^3 \left(c^4-22 c^3+41 c^2-22 c+2\right)\no\\
    &\quad+b^2 c \left(4 c^3-29 c^2+19 c-4\right)+2 b c^2 \left(5 c^2-9 c+4\right)+c^3 (3 c+2)\Bigr)\no\\
    &\quad+2 a (b-1)^3 b (b+1) (c-2) (c-1)^2 c (b-c)+4 (b-1)^3 b (b+1) (c-1)^3 c (b-c)\Bigr)\,,
\end{align}
\begin{align}
 \sigma_0^{3,3}&=4 a^2 c \Bigl(a^6 (b+1)^2 (b-c)^2 \left(b^2 \left(c^2-12 c-4\right)+b \left(8 c^2+20 c+2\right)-c (11 c+4)\right)\no\\
 &\quad-2 a^5 (b+1)^2 (b-c)^2 \left(b^2 \left(4 c^2+7 c+4\right)-3 b \left(3 c^2+6 c+1\right)+3 c (2 c+3)\right)\no\\
 &\quad-a^4 \left(b^2-1\right) (c-1) \Bigl(b^4 \left(5 c^2-20 c+3\right)+b^3 \left(-5 c^3+34 c^2+5 c-2\right)\no\\
 &\quad-b^2 \left(14 c^3+9 c^2-18 c+3\right)+b c \left(c^2-32 c-1\right)+4 c^2 (4 c+1)\Bigr)\no\\
 &\quad+2 a^3 \left(b^2-1\right) \Bigl(b^4 \left(8 c^3+4 c^2+3\right)-b^3 \left(8 c^4+6 c^3+23 c^2-8 c+1\right)\no\\
 &\quad+b^2 \left(2 c^4+13 c^3-11 c^2-c-3\right)+b c \left(10 c^3+7 c^2+14 c-1\right)+c^2 \left(-7 c^2-12 c+4\right)\Bigr)\no\\
 &\quad+a^2 (b-1)^2 \Bigl(b^4 \left(8 c^4-28 c^3+51 c^2-28 c+12\right)+2 b^3 \left(6 c^4-11 c^3+2 c+3\right)\no\\
 &\quad+b^2 \left(3 c^4-2 c^3-31 c^2-2 c+2\right)-2 b c \left(9 c^3-24 c^2+14 c+1\right)+3 c^3 (8-3 c)\Bigr)\no\\
 &\quad-2 a (b-1)^2 (c-1) \Bigl(b^4 \left(8 c^2-8 c+5\right)+4 b^3 (c-1)^2 (2 c-1)-b^2 \left(4 c^3+2 c^2+3 c+1\right)-3 b (c-1)^2 c+c^2 (c+4)\Bigr)\no\\
 &\quad+(b-1)^3 (c-1)^2 \left(b^3 \left(8 c^2-7 c+5\right)+b^2 \left(-c^2+4 c+1\right)+3 b c (1-3 c)-4 c^2\right)\Bigr)\,,\\
  \sigma_0^{2,3}&=4 a \Bigl(a^7 (b+1)^2 (b-c)^2 \left(b^2 \left(c^3-5 c^2-18 c-3\right)+b c \left(2 c^2+35 c+13\right)-c^2 (14 c+11)\right)\no\\
  &\quad-a^6 (b+1)^2 (b-c)^2 \left(b^2 \left(c^3+23 c^2+21 c+5\right)-b \left(13 c^3+53 c^2+33 c+1\right)+2 c \left(6 c^2+17 c+2\right)\right)\no\\
  &\quad-a^5 \left(b^2-1\right) (c-1) c \Bigl(b^4 \left(2 c^2-7 c-33\right)+b^3 \left(-2 c^3+15 c^2+77 c-12\right)+b^2 \left(-8 c^3-62 c^2+61 c+7\right)\no\\
  &\quad+b \left(18 c^3-75 c^2-19 c-2\right)+2 c \left(13 c^2+6 c+1\right)\Bigr)\no\\
  &\quad+a^4 \left(b^2-1\right) \Bigl(b^4 \left(3 c^4+57 c^3-11 c^2+5 c-4\right)-b^3 \left(3 c^5+73 c^4+15 c^3+35 c^2-28 c+2\right)\no\\
  &\quad+b^2 c \left(16 c^4+21 c^3-23 c^2-19 c+5\right)+b c \left(5 c^4+69 c^3+37 c^2-7 c-4\right)+2 c^2 \left(-8 c^3-21 c^2+2 c+2\right)\Bigr)\no\\
  &\quad+a^3 (b-1)^2 \Bigl(b^4 \left(10 c^4-25 c^3+63 c^2-28 c+5\right)+b^3 \left(14 c^5-14 c^4-30 c^3+37 c^2-9 c+2\right)\no\\
  &\quad+b^2 c \left(-9 c^4+65 c^3-148 c^2+50 c-8\right)-b c^2 \left(26 c^3-61 c^2+33 c+2\right)+c^3 \left(-18 c^2+35 c+8\right)\Bigr)\no\\
  &\quad-a^2 (b-1)^2 (c-1) \Bigl(b^4 \left(4 c^4+12 c^3+5 c^2-2 c+1\right)+b^3 \left(-20 c^4+62 c^3-75 c^2+36 c-3\right)\no\\
  &\quad+5 b^2 c \left(c^3-12 c^2+5 c-2\right)+b c \left(c^3+2 c^2+c-4\right)+4 c^2 \left(c^2+3 c+1\right)\Bigr)\no\\
  &\quad+a (b-1)^3 (c-1)^2 \left(b^3 \left(12 c^2-9 c+2\right)+b^2 c \left(4 c^2+c+3\right)+b c \left(-8 c^2+c+2\right)-2 c^2 (3 c+1)\right)\no\\
  &\quad+(b-1)^3 b (c-1)^3 c \left(b^2 (4 c-3)-b c+b-c\right)\Bigr)\,,\\
 \sigma_0^{1,3}&=2 a \Bigl(a^7 \left(-(b+1)^2\right) (b-c)^2 \left(b^2 \left(c^3-14 c^2+40 c+8\right)+b c \left(9 c^2-50 c-29\right)+c^2 (17 c+18)\right)\no\\
 &\quad+a^6 (b+1)^2 (b-c)^2 \left(b^2 \left(7 c^3-13 c^2-60 c-4\right)+b \left(5 c^3+66 c^2+57 c+12\right)-c \left(15 c^2+39 c+16\right)\right)\no\\
 &\quad+a^5 \left(b^2-1\right) (c-1) \Bigl(b^4 \left(5 c^3-24 c^2+81 c+12\right)+b^3 \left(-5 c^4+36 c^3-199 c^2+16 c+8\right)\no\\
 &\quad-2 b^2 \left(6 c^4-86 c^3+55 c^2+25 c+2\right)+6 b c \left(-9 c^3+20 c^2+11 c+2\right)-2 c^2 \left(19 c^2+12 c+4\right)\Bigr)\no\\
 &\quad+a^4 \left(b^2-1\right) \Bigl(b^4 \left(c^4+24 c^3+131 c^2-90 c+4\right)-b^3 \left(c^5+57 c^4+178 c^3-117 c^2+29 c-8\right)\no\\
 &\quad+b^2 \left(33 c^5+66 c^4-113 c^3-34 c^2+44 c+4\right)+b c \left(-19 c^4+116 c^3+91 c^2-32 c-16\right)-2 c^2 \left(15 c^3+16 c^2+10 c-6\right)\Bigr)\no\\
 &\quad+a^3 (b-1)^2 \Bigl(b^4 \left(15 c^4-58 c^3+117 c^2-31 c-8\right)+b^3 c \left(-15 c^4+141 c^3-326 c^2+291 c-91\right)\no\\
 &\quad+b^2 c \left(-35 c^4+186 c^3-315 c^2+106 c-12\right)+b c \left(-73 c^4+212 c^3-215 c^2+84 c-8\right)+c^2 \left(-29 c^3+48 c^2+8 c+8\right)\Bigr)\no\\
 &\quad-a^2 (b-1)^2 (c-1) \Bigl(b^4 c \left(19 c^2+56 c-40\right)+b^3 \left(-3 c^4-60 c^3+119 c^2-68 c+12\right)\no\\
 &\quad+b^2 \left(-28 c^4+7 c^3-101 c^2+56 c-4\right)+2 b c \left(21 c^3-37 c^2+22 c-6\right)+c^2 \left(19 c^2+16\right)\Bigr)\no\\
 &\quad+2 a (b-1)^3 (c-1)^2 \left(b^3 \left(5 c^2-5 c-2\right)+b^2 \left(-5 c^3+22 c^2-15 c+2\right)+b c \left(-9 c^2+13 c-2\right)-4 c^3\right)\no\\
 &\quad+4 (b-1)^3 (c-1)^3 c \left(2 b^3+b^2 (2 c-1)-3 b c+b-c\right)\Bigr)\,,\\
  \sigma_0^{0,3}&=-2 a \Bigl(a^7 (b+1)^2 (b-c)^2 \left(b^2 \left(c^3-9 c^2+14 c+3\right)+b c \left(5 c^2-13 c-10\right)+c^2 (4 c+5)\right)\no\\
  &\quad-a^6 (b+1)^2 (b-c)^2 \left(b^2 \left(14 c^2-37 c+5\right)+b \left(-3 c^3+22 c^2+7 c+10\right)-c \left(5 c^2+5 c+8\right)\right)\no\\
  &\quad+a^5 \left(b^2-1\right) (c-1)
  \bigl(b^4 \left(5 c^2-17 c-12\right)+b^3 \left(-15 c^3+63 c^2+6 c-8\right)\no\\
  &\qquad+b^2 \left(10 c^4-67 c^3+25 c^2+30 c+4\right)+b c \left(21 c^3-29 c^2-30 c-8\right)+2 c^2 \left(5 c^2+4 c+2\right)\bigr)\no\\
  &\quad+a^4 \left(b^2-1\right) \bigl(b^4 \left(2 c^4-51 c^2+26 c+5\right)-b^3 \left(2 c^5+5 c^4-134 c^3+145 c^2-66 c+12\right)\no\\
  &\qquad+b^2 \left(5 c^5-104 c^4+183 c^3-106 c^2+42 c-20\right)+b c \left(21 c^4-76 c^3+37 c^2-42 c+24\right)+2 c^2 \left(6 c^3-c^2+6 c-2\right)\bigr)\no\\
  &\quad-a^3 (b-1)^2 \bigl(b^4 \left(c^4-4 c^3+2 c^2+27 c-17\right)-b^3 \left(c^5-21 c^4+36 c^3+10 c^2-54 c+28\right)\no\\
  &\qquad-b^2 \left(17 c^5-69 c^4+76 c^3+18 c^2-32 c+8\right)+b c \left(-19 c^4+37 c^3+2 c^2-44 c+24\right)+c^2 \left(-10 c^3+19 c^2-8 c+8\right)\bigr)\no\\
  &\quad+a^2 (b-1)^2 (c-1) \bigl(b^4 \left(2 c^3+24 c^2-5 c-10\right)+b^3 \left(-2 c^4-37 c^3+74 c^2-37 c+2\right)\no\\
  &\qquad+b^2 \left(21 c^4-107 c^3+108 c^2-64 c+20\right)+2 b c \left(11 c^3-16 c^2+3 c+2\right)+c^2 \left(11 c^2-8 c+8\right)\bigr)\no\\
  &\quad-a (b-1)^3 (c-1)^2 \left(b^3 \left(c^2+2 c-8\right)-b^2 \left(c^3+c^2-14 c+12\right)-b c \left(c^2+2 c-8\right)-4 (c-1) c^2\right)\no\\
  &\quad-2 (b-1)^3 (c-1)^3 c \left(3 b^3+b^2 (5-3 c)-b (c+2)-2 c\right)\Bigr)\,,
\end{align}
\begin{align}
 \sigma_0^{2,2}&=4 \Bigl( a^8(b+1)^2 (b-c)^2 \left(\left(c^3-18 c-18\right) b^2-2 c \left(c^2-15 c-21\right) b-c^2 (11 c+24)\right)\no\\
 &\quad+2 a^7(b+1)^2 (b-c)^2 \left(\left(c^3-8 c^2-17 c-11\right) b^2+\left(4 c^3+21 c^2+42 c+3\right) b-c \left(2 c^2+27 c+6\right)\right) \no\\
 &\quad- a^6\left(b^2-1\right) (c-1) \Bigl(\left(c^3-47 c-26\right) b^4+\left(-c^4+2 c^3+123 c^2+20 c-12\right) b^3\no\\
 &\quad+\left(-2 c^4-109 c^3+62 c^2+59 c+2\right) b^2+3 c \left(11 c^3-28 c^2-25 c-2\right) b+4 c^2 \left(7 c^2+7 c+1\right)\Bigr)\no\\
 &\quad-2 a^5 \left(b^2-1\right) \Bigl(\left(c^4-31 c^3-13 c^2-2 c+10\right) b^4-\left(c^5-49 c^4+4 c^3-69 c^2+49 c-6\right) b^3\no\\
 &\quad-2 \left(9 c^5-8 c^4+26 c^3-46 c^2+18 c+1\right) b^2+c \left(c^4-18 c^3-89 c^2+30 c+6\right) b+c^2 \left(3 c^3+36 c^2-4\right)\Bigr)\no\\
 &\quad+ a^4(b-1)^2 \Bigl(\left(8 c^4-25 c^3+68 c^2-6 c-10\right) b^4+\left(-8 c^5+114 c^4-286 c^3+262 c^2-86 c+4\right) b^3\no\\
 &\quad+\left(-17 c^5+126 c^4-233 c^3+44 c^2+14 c-4\right) b^2+2 c \left(-26 c^4+73 c^3-64 c^2+15 c+2\right) b+c^3 \left(-25 c^2+36 c+24\right)\Bigr)\no\\
 &\quad-2a^3 (b-1)^2 (c-1) \Bigl(\left(22 c^3-15 c^2+15 c-7\right) b^4+\left(-10 c^4+16 c^3-7 c^2-2 c+3\right) b^3\no\\
 &\quad+\left(-25 c^4+54 c^3-98 c^2+41 c-2\right) b^2+2 c \left(5 c^3-7 c^2+3 c-1\right) b+c^2 \left(c^2+10 c+4\right)\Bigr) \no\\
 &\quad-a^2(b-1)^3 (c-1)^2 \left(\left(8 c^3-25 c^2+15 c+2\right) b^3+\left(17 c^3-56 c^2+29 c-2\right) b^2+c \left(9 c^2-7 c-2\right) b+4 c^2 (2 c+1)\right) \no\\
 &\quad+6 a c b(c-1)^3(b-1)^3  \left(b^2+3 (c-1) b-c\right) -8 (b-1)^4 b^2 (c-1)^4 c\Bigr)\,,\\
 \sigma_0^{1,2}&=-2 \Bigl(\Bigl(\left(c^3-10 c^2+12 c+42\right) a^8+\left(-4 c^3-2 c^2+51 c+45\right) a^7+\left(-3 c^4+17 c^3-61 c^2-33 c+80\right) a^6\no\\
 &\quad+\left(-18 c^3-143 c^2+28 c+43\right) a^5+\left(-15 c^4+54 c^3-83 c^2-47 c+46\right) a^4+\left(16 c^4+76 c^3-149 c^2+51 c+6\right) a^3\no\\
 &\quad-2 (c-1)^2 \left(4 c^2-3 c-6\right) a^2-2 (c-1)^3 \left(4 c^2+c+4\right) a+4 (c-1)^4\Bigr) b^6\no\\
 &\quad+\Bigl(\left(-2 c^4+28 c^3-59 c^2-141 c+84\right) a^8+\left(8 c^4-11 c^3-128 c^2-109 c+60\right) a^7\no\\
 &\quad+\left(3 c^5-33 c^4+181 c^3+3 c^2-226 c+72\right) a^6+\left(45 c^4+198 c^3-11 c^2-64 c+12\right) a^5\no\\
 &\quad+c \left(15 c^4-105 c^3+131 c^2+112 c-63\right) a^4-\left(16 c^5+13 c^4+159 c^3-361 c^2+203 c-30\right) a^3\no\\
 &\quad+10 (c-1)^2 \left(4 c^3-7 c^2+2 c-2\right) a^2+2 (c-1)^3 \left(17 c^2-12 c+16\right) a+4 (c-1)^4 (5 c-4)\Bigr) b^5\no\\
 &\quad+ b^4\Bigl(\left(c^5-26 c^4+103 c^3+153 c^2-318 c+42\right) a^8+\left(-4 c^5+28 c^4+88 c^3+92 c^2-279 c-15\right) a^7\no\\
 &\quad+\left(16 c^5-163 c^4+75 c^3+329 c^2-169 c-88\right) a^6-\left(27 c^5+34 c^4+99 c^3-495 c^2+174 c+71\right) a^5\no\\
 &\quad+\left(51 c^5-135 c^4+150 c^3-154 c^2+267 c-134\right) a^4+\left(9 c^5-48 c^4+109 c^3-269 c^2+313 c-114\right) a^3\no\\
 &\quad-4 (c-1)^2 \left(16 c^3-39 c^2+15 c+3\right) a^2-16 (c-1)^4 (5 c-3) a-8 (c-1)^4 (10 c-3)\Bigr)\no\\
 &\quad+\Bigl(c \left(8 c^4-77 c^3-29 c^2+443 c-165\right) a^8-\left(15 c^5-4 c^4+38 c^3-466 c^2+27 c+30\right) a^7\no\\
 &\quad+\left(43 c^5-59 c^4-329 c^3+179 c^2+238 c-72\right) a^6+\left(-21 c^5+43 c^4-615 c^3+157 c^2+88 c-12\right) a^5\no\\
 &\quad
 +\left(-57 c^5+418 c^4-788 c^3+396 c^2-233 c+84\right) a^4+\left(-4 c^5+301 c^4-527 c^3+461 c^2-345 c+114\right) a^3\no\\
 &\quad-4 (c-1)^2 \left(7 c^3-3 c^2-9\right) a^2+4 (c-1)^3 \left(32 c^2-55 c+8\right) a+8 (c-1)^4 (15 c-2)\Bigr) b^3\no\\
 &\quad+(a+1) \Bigl(c^2 \left(21 c^3-47 c^2-260 c+241\right) a^7+c \left(-36 c^4+58 c^3-72 c^2-87 c+92\right) a^6\no\\
 &\quad+\left(50 c^5+134 c^4-72 c^3-185 c^2+110 c+8\right) a^5+\left(-11 c^5-14 c^4+177 c^3-163 c^2+36 c+20\right) a^4\no\\
 &\quad+\left(-46 c^5+7 c^4+163 c^3-224 c^2+116 c-16\right) a^3+2 (c-1)^2 \left(56 c^3-137 c^2+76 c-10\right) a^2\no\\
 &\quad-2 (c-1)^3 \left(22 c^2-39 c+2\right) a-4 (c-1)^4 (20 c-1)\Bigr) b^2\no\\
 &\quad+(a+1)^2 c \Bigl(c^2 \left(22 c^2+43 c-155\right) a^6-(c-1)^2 c (45 c+94) a^5\no\\
 &\quad
 +\left(22 c^4+57 c^3+17 c^2+6 c-12\right) a^4+2 c \left(11 c^3-99 c^2+120 c-32\right) a^3\no\\
 &\quad
 -2 (c-1)^2 \left(30 c^2-67 c+32\right) a^2+2 (c-1)^3 (11 c-2) a+20 (c-1)^4\Bigr) b\no\\
 &\quad+(a-1) a (a+1)^3 c^2 \left(c^2 (8 c+37) a^3+c \left(-13 c^2-19 c+32\right) a^2-4 (c-1)^3 a+12 (c-1)^3\right)\Bigr)\,,\\
 \sigma_0^{0,2}&=-2 \Bigl(\Bigl(\left(c^3-5 c^2+15\right) a^8+\left(c^3-13 c^2+22 c+12\right) a^7\no\\
 &\quad+\left(3 c^3-6 c^2-31 c+34\right) a^6+\left(c^4+4 c^3-49 c^2+2 c+20\right) a^5\no\\
 &\quad+\left(c^4-12 c^3+40 c^2-75 c+35\right) a^4+\left(5 c^4+5 c^3+12 c^2-46 c+24\right) a^3\no\\
 &\quad+5 \left(c^2-3 c+2\right)^2 a^2-8 (c-1)^3 a+4 (c-1)^4\Bigr) b^6\no\\
 &\quad
 -\Bigl(2 \left(c^4-7 c^3+4 c^2+28 c-15\right) a^8+\left(2 c^4-29 c^3+80 c^2-5 c-4\right) a^7\no\\
 &\quad+\left(9 c^4-20 c^3-95 c^2+154 c-48\right) a^6+\left(c^5+7 c^4-100 c^3+33 c^2+31 c-16\right) a^5\no\\
 &\quad+\left(c^5-3 c^4+47 c^3-123 c^2+66 c-10\right) a^4+\left(5 c^5+15 c^4-25 c^3+39 c^2-62 c+28\right) a^3\no\\
 &\quad+(c-1)^2 \left(5 c^3-18 c^2-4 c+24\right) a^2+4 (c-1)^3 \left(2 c^2-c-4\right) a+4 (c-1)^4 (c+2)\Bigr) b^5\no\\
 &\quad
 +\Bigl(\left(c^5-13 c^4+22 c^3+76 c^2-112 c+15\right) a^8+\left(c^5-19 c^4+100 c^3-74 c^2-2 c-28\right) a^7\no\\
 &\quad+\left(6 c^5-23 c^4-79 c^3+207 c^2-73 c-38\right) a^6+\left(3 c^5-59 c^4+55 c^3+39 c^2+48 c-64\right) a^5\no\\
 &\quad+\left(9 c^5+6 c^4-47 c^3-9 c^2+125 c-73\right) a^4+\left(10 c^5-24 c^4-21 c^3+33 c^2+30 c-28\right) a^3\no\\
 &\quad+2 (c-1)^2 \left(9 c^3-32 c^2+32 c-16\right) a^2+4 (c-1)^3 \left(9 c^2-8 c+4\right) a+8 (c-1)^4 (2 c-1)\Bigr) b^4\no\\
 &\quad
 +\Bigl(4 c \left(c^4-5 c^3-10 c^2+39 c-14\right) a^8+\left(3 c^5-48 c^4+92 c^3-8 c^2+69 c-20\right) a^7\no\\
 &\quad+\left(9 c^5+15 c^4-115 c^3-19 c^2+158 c-48\right) a^6+\left(8 c^5-35 c^4-71 c^3-61 c^2+95 c-24\right) a^5\no\\
 &\quad+\left(c^5-29 c^4+191 c^3-379 c^2+204 c-32\right) a^4+\left(19 c^5-21 c^4+115 c^3-187 c^2+70 c+4\right) a^3\no\\
 &\quad-2 (c-1)^2 \left(12 c^3-21 c^2+16 c-20\right) a^2-8 (c-1)^3 \left(7 c^2-10 c+8\right) a-8 (c-1)^4 (3 c-4)\Bigr) b^3\no\\
 &\quad
 +(a+1) \Bigl(c^2 \left(6 c^3+c^2-95 c+77\right) a^7+c \left(-43 c^3+104 c^2-126 c+54\right) a^6\no\\
 &\quad+\left(80 c^4-48 c^3-75 c^2+50 c+4\right) a^5+\left(11 c^5-30 c^4+17 c^3+65 c^2-92 c+40\right) a^4\no\\
 &\quad+\left(-31 c^5+65 c^4-62 c^3+68 c^2-68 c+28\right) a^3-(c-1)^2 \left(2 c^3-63 c^2+100 c-40\right) a^2\no\\
 &\quad+4 (c-1)^4 (4 c-7) a+4 (c-1)^4 (4 c-7)\Bigr) b^2\no\\
 &\quad
 +(a+1)^2 \Bigl(2 c^3 \left(2 c^2+10 c-23\right) a^6-(c-1)^2 c^2 (c+48) a^5\no\\
 &\quad+c \left(-11 c^4+66 c^3-49 c^2+20 c-4\right) a^4+2 \left(8 c^5-22 c^4-13 c^3+51 c^2-28 c+4\right) a^3\no\\
 &\quad-(c-1)^2 \left(15 c^3+36 c^2-56 c+24\right) a^2+4 (c-1)^3 c (2 c+5) a-4 (c-2) (c-1)^4\Bigr) b\no\\
 &\quad
 +(a-1) a (a+1)^3 c \left(\left(a^3+a^2-8 a+4\right) c^4+\left(10 a^3-15 a^2+16 a-4\right) c^3+2 \left(7 a^2-4 a-6\right) c^2+20 c-8\right)\Bigr)\,,
\end{align}

\begin{align}
\sigma_0^{1,1}&=b^6\Bigl(\left(c^3-8 c^2+48 c-96\right) a^8+2 \left(3 c^3-2 c^2-56\right) a^7+8 \left(2 c^3-9 c^2+33 c-26\right) a^6\no\\
 &\quad+\left(-4 c^4+30 c^3+84 c^2+144 c-144\right) a^5+\left(4 c^4+27 c^3-132 c^2+300 c-144\right) a^4\no\\
 &\quad-4 \left(c^4+31 c^3-42 c^2+2 c+8\right) a^3+4 (c-1)^2 \left(5 c^2-3 c-4\right) a^2+32 (c-1)^3 (2 c-1) a+16 (c-1)^4\Bigr) \no\\
 &\quad
 -2  b^5\Bigl(\left(c^4-10 c^3+76 c^2-218 c+96\right) a^8+2 \left(3 c^4-9 c^3+8 c^2-93 c+36\right) a^7\no\\
 &\quad+4 \left(c^4-11 c^3+83 c^2-101 c+28\right) a^6+\left(-2 c^5+20 c^4+96 c^3+74 c^2-94 c+16\right) a^5\no\\
 &\quad+\left(2 c^5-15 c^4-40 c^3+186 c^2-86 c+8\right) a^4-2 \left(c^5+33 c^4-59 c^3+57 c^2-44 c+12\right) a^3\no\\
 &\quad+2 (c-1)^2 \left(5 c^3+15 c^2-22 c+8\right) a^2+8 (c-1)^3 (13 c-10) a+8 (c-1)^4 (c+5)\Bigr)\no\\
 &\quad
 +\Bigl(\left(c^5-16 c^4+166 c^3-728 c^2+728 c-96\right) a^8+2 \left(3 c^5-30 c^4+50 c^3-252 c^2+260 c+24\right) a^7\no\\
 &\quad-4 \left(2 c^5+3 c^4-137 c^3+263 c^2-83 c-48\right) a^6+2 \left(5 c^5+26 c^4+80 c^3-242 c^2-52 c+128\right) a^5\no\\
 &\quad+\left(-57 c^5+28 c^4+246 c^3-204 c^2-324 c+256\right) a^4-4 \left(2 c^5+48 c^4-281 c^3+415 c^2-204 c+20\right) a^3\no\\
 &\quad-8 (c-1)^2 \left(7 c^3-29 c^2+26 c-20\right) a^2-16 (c-1)^3 \left(c^2-16 c+20\right) a+16 (c-1)^4 (c+10)\Bigr) b^4\no\\
 &\quad
 +4 \Bigl(c \left(c^4-17 c^3+134 c^2-258 c+85\right) a^8+\left(7 c^5-26 c^4+88 c^3-180 c^2-19 c+20\right) a^7\no\\
 &\quad-\left(c^5+37 c^4-143 c^3-29 c^2+190 c-56\right) a^6+\left(14 c^5-41 c^4+201 c^3-25 c^2-47 c+8\right) a^5\no\\
 &\quad+\left(6 c^5+4 c^4-179 c^3+423 c^2-239 c+40\right) a^4+\left(17 c^5-81 c^4-109 c^3+393 c^2-280 c+60\right) a^3\no\\
 &\quad+2 (c-1)^2 \left(26 c^3-51 c^2+27 c-16\right) a^2+8 (c-1)^3 \left(3 c^2-8 c+10\right) a+8 (c-1)^4 (3 c-5)\Bigr) b^3\no\\
 &\quad
 +(a+1) \Bigl(c^2 \left(6 c^3-152 c^2+649 c-448\right) a^7+c \left(30 c^4+32 c^3-195 c^2+412 c-224\right) a^6\no\\
 &\quad-\left(30 c^5+156 c^4+193 c^3-680 c^2+372 c-16\right) a^5+\left(38 c^5-148 c^4+243 c^3-392 c^2+332 c-128\right) a^4\no\\
 &\quad+4 \left(15 c^5-23 c^4+31 c^3-82 c^2+103 c-44\right) a^3-4 (c-1)^2 \left(30 c^3-29 c^2-7 c+24\right) a^2\no\\
 &\quad+16 (c-1)^4 (2 c+5) a-16 (c-1)^4 (14 c-5)\Bigr) b^2\no\\
 &\quad
 +2 (a+1)^2 \Bigl(c^3 \left(2 c^2-77 c+130\right) a^6+2 (c-1)^2 c^2 (c+52) a^5\no\\
 &\quad-c \left(4 c^4+45 c^3+48 c^2-66 c+24\right) a^4-4 c \left(6 c^4-23 c^3+c^2+28 c-12\right) a^3\no\\
 &\quad+2 (c-1)^2 \left(21 c^3+9 c^2-8 c+12\right) a^2-8 (c-1)^3 c (12 c-5) a+8 (c-1)^4 (11 c-1)\Bigr) b\no\\
 &\quad
 + c (a-1) (a+1)^3\Bigl((a-2)^2 \left(a^2+12\right) c^4-4 \left(14 a^4-17 a^3-36 a+48\right) c^3\no\\
 &\quad-16 \left(4 a^3+3 a^2+9 a-18\right) c^2+16 \left(2 a^2+3 a-12\right) c+48\Bigr)\,,\\
\sigma_0^{0,1}&=a^8 (b+1)^2 (b-c)^2 \left(b^2 \left(c^3-8 c^2+28 c-34\right)+b c \left(3 c^2-23 c+46\right)+c^2 (2 c-15)\right)\no\\
 &\quad
 +2 a^7 (b-1) (b+1)^2 (b-c)^2 \left(b \left(c^3-2 c^2+8 c-20\right)+c \left(2 c^2-c+12\right)\right)\no\\
 &\quad
 +2 a^6 \left(b^2-1\right) (c-1) \Bigl(b^4 \left(5 c^2-22 c+40\right)+b^3 \left(-10 c^3+63 c^2-144 c+52\right)\no\\
 &\quad+b^2 \left(5 c^4-51 c^3+161 c^2-134 c+12\right)+b c \left(10 c^3-61 c^2+114 c-24\right)+4 c^2 \left(c^2-8 c+3\right)\Bigr)\no\\
 &\quad
 +2 a^5 (b-1)^2 (b+1) \Bigl(b^3 \left(3 c^3-4 c^2+46 c-32\right)+b^2 \left(-10 c^4+24 c^3-103 c^2+124 c-48\right)\no\\
 &\quad+b \left(7 c^5-26 c^4+50 c^3-100 c^2+72 c-16\right)+c \left(6 c^4+7 c^3+8 c^2-24 c+16\right)\Bigr)\no\\
 &\quad
 -a^4 (b-1)^2 \Bigl(b^4 \left(2 c^4-41 c^3+128 c^2-172 c+70\right)+b^3 \left(-2 c^5+74 c^4-361 c^3+687 c^2-590 c+192\right)\no\\
 &\quad+b^2 \left(-33 c^5+272 c^4-808 c^3+1091 c^2-632 c+136\right)+b \left(-39 c^5+307 c^4-764 c^3+808 c^2-344 c+32\right)\no\\
 &\quad-14 c^5+97 c^4-176 c^3+96 c^2-16\Bigr)\no\\
 &\quad
 +4 a^3 (b-1)^3 (c-1) \Bigl(-b^3 \left(c^3-2 c^2+15 c-10\right)+b^2 \left(c^4-8 c^3+13 c^2-38 c+28\right)\no\\
 &\quad+2 b \left(3 c^4+c^2-10 c+8\right)+2 \left(c^4+3 c^3+6 c^2-12 c+4\right)\Bigr)\no\\
 &\quad
 -2 a^2 (b-1)^3 (c-1)^2 \Bigl(b^3 \left(5 c^2-22 c+16\right)+b^2 \left(-5 c^3+35 c^2-50 c+36\right)\no\\
 &\quad+b \left(-13 c^3+70 c^2-88 c+32\right)-4 c \left(c^2-5 c+8\right)\Bigr)\no\\
 &\quad
 -8 a (b-1)^4 (c-1)^3 \left(b^2 (c-2)-b c (c+4)-4\right)-8 (b-1)^4 (c-1)^4 \left(b^2-b c+2\right)\,,\\
\sigma_0^{0,0}&=\left(a^4 (b+1) \Bigl(b^2 (c-2)-b (c-3) c-c^2\right)\no\\
&\quad+a^2 (b-1) \left(b^2 (3 c-2)+b \left(-3 c^2+7 c-4\right)-5 c^2+8 c-4\right)-4 (b-1)^2 (c-1)^2\Bigr)\no\\ 
&\quad\times\Bigl(-(a+1)^2 c \left(a^2 (c-4) c-4 a \left(c^2-3 c+2\right)+4 (c-1)^2\right)\no\\
&\qquad\quad-(a+1) b \left(a^3 c \left(2 c^2-9 c+10\right)+a^2 \left(-2 c^3+11 c^2-14 c+8\right)-4 a \left(2 c^2-3 c+1\right)-4 (c-1)^2 (2 c+1)\right)\no\\
&\qquad\quad+b^3 \left(a^4 \left(c^2-4 c+6\right)+2 a^3 \left(c^2-2 c+4\right)+a^2 \left(5 c^2-12 c+10\right)+4 a \left(c^2-3 c+2\right)+4 (c-1)^2\right)\no\\
&\qquad\quad-b^2 \Bigl(a^4 \left(c^3-6 c^2+14 c-6\right)+2 a^3 c \left(c^2-2 c+4\right)\no\\
&\qquad\quad+a^2 \left(5 c^3-10 c^2+2 c+6\right)+4 a \left(c^3+c^2-6 c+4\right)+4 \left(c^3-3 c+2\right)\Bigr)\Bigr)\,.
\end{align}

\end{document}